\documentclass[12pt, draftclsnofoot, onecolumn]{IEEEtran}

\usepackage{epstopdf}
\usepackage{multirow}
\usepackage{amsmath,epsfig,verbatim,amsopn,subfigure,color}
\usepackage{cite,xspace}
\usepackage{array}
\usepackage[ruled,lined,boxed,commentsnumbered]{algorithm2e}
\usepackage{array}
\usepackage{multirow,array}%
\usepackage{multicol}%
\usepackage{setspace}
\usepackage{here}
\usepackage{amssymb}
\usepackage{dsfont}
\usepackage{amsthm}
\usepackage{subfigure}
\usepackage{soul}

\usepackage{breqn}

\usepackage[center]{caption}
\usepackage{tikz}
\usetikzlibrary{matrix,fit}
\usetikzlibrary{shapes,arrows}

\usepackage{ graphicx,  bm,  bbding,pifont}
\usepackage{epsf, epsfig}
\usepackage{graphics}
\usepackage{times,relsize}
\usepackage{fancybox,calc}
\usepackage{graphics, color, psfrag,amsfonts}
\usetikzlibrary{decorations.markings}
\usepackage{sidecap}
\usetikzlibrary{shapes,arrows}
\usetikzlibrary{decorations.pathreplacing}
\tikzstyle{decision} = [diamond, draw, fill=white!20,
    text width= 2em, text centered, node distance=1cm, inner sep=2.1pt]
\tikzstyle{block} = [rectangle, draw, fill=white!20,
    text width=2em, text centered, rounded corners, minimum height=2.1em]
\tikzstyle{new_block} = [rectangle, draw, fill=gray!30,
    text width=2em, text centered, rounded corners, minimum height=2.1em]
\tikzstyle{block2} = [rectangle, draw, fill=white!20,
    text width=2em, text centered, rounded corners, minimum height=2.7em]
\tikzstyle{line} = [draw, -latex']
\tikzstyle{state}=[circle, draw, fill=white!10,
    text width=2em, text badly centered, inner sep=2.5pt,minimum height=1.5em]
\tikzstyle{cloud} = [ellipse,fill=gray!35, node distance=3cm,
    minimum height=2em]

\ifCLASSOPTIONonecolumn

\else
    
\fi

\ifCLASSOPTIONonecolumn
\usepackage[
top    = 0.85in,
bottom = 0.85in,
left   = 2.3cm,
right  = 2.3cm
    ]{geometry}
    \else
    \fi

\begin{document}
\title{
Network Coding for Video  Distortion Reduction  in Device-to-Device Communications}

\author{\authorblockN{Mohammad S. Karim, Ëœ\IEEEmembership{Student Member, ËœIEEE,
}  Sameh Sorour, Ëœ\IEEEmembership{Member, ËœIEEE,
}  and  Parastoo Sadeghi, Ëœ\IEEEmembership{Senior Member, ËœIEEE
}
 }
\thanks{M. S. Karim and P. Sadeghi   are with the  Australian National University,
Australia  (\{mohammad.karim, parastoo.sadeghi\}@anu.edu.au). S. Sorour is with the King Fahd University of Petroleum and Minerals, Saudi Arabia (samehsorour@kfupm.edu.sa).}
}
\maketitle

\begin{abstract}

%

In this paper, we study the problem of distributing  a real-time  video sequence  to a group of partially connected cooperative wireless devices  using instantly decodable network coding  (IDNC). In such a scenario, the coding conflicts occur to service multiple devices with an immediately decodable  packet and the transmission conflicts occur from simultaneous  transmissions of multiple devices. To avoid  these conflicts, we introduce a novel  IDNC graph that represents  all feasible coding and transmission conflict-free decisions  in one unified framework. Moreover, a real-time video sequence has a hard  deadline and unequal importance of video packets. Using these video characteristics and the new  IDNC graph, we formulate the problem of minimizing the mean video distortion before the deadline as a finite horizon Markov decision process (MDP) problem. However, the backward induction algorithm that finds the optimal policy of the  MDP formulation  has high modelling and computational complexities. To reduce  these complexities,  we further design a two-stage maximal independent set selection  algorithm, which can efficiently reduce  the mean video distortion before the deadline. Simulation results over a real video sequence show that our proposed IDNC  algorithms improve the received video quality compared to the existing IDNC algorithms.
\end{abstract}



\vspace{-1mm}
\section{Introduction} \label{introduction}
There is a sharp increase in the demand for high quality content  over wireless networks. The simultaneous  increase in the popularity of smart devices with improved computational, storage and connectivity capabilities is expected to play an important role in addressing the increased  throughput demand of  wireless networks. This leads to a heterogenous network architecture, where smart devices  use two wireless interfaces simultaneously. One interface communicates with the central station using a \emph{long-range wireless technology}, e.g., GSM, WiMAX or LTE, and the other interface communicates with other smart devices using a \emph{short-range wireless technology}, e.g., Bluetooth or 802.11 adhoc mode. The usage of a short-range wireless technology  has numerous practical  advantages  \cite{al2013energy,khamfroush2014coded,el2011interplay,hernandez2014throughput,abedini2013realtime}. First, it  offloads the central station  to serve additional  devices and increase the  throughput of the  network.  Second, it increases the coverage zone  of the network as devices can communicate to other devices   via intermediate devices.  Third, it reduces the cost associated with the deployment of new infrastructure required for the  growing network size and devices' throughput demand.  Finally,  short-range  channels  provide  more reliable delivery of the packets compared to the long-range channels due to  small distances between the  devices.

In this paper, we are interested in distributing a real-time video sequence  to a group of partially connected cooperative  wireless devices. Such a real-time video sequence has two distinct characteristics \cite{nguyen2007multimedia,seferoglu2009video}.  First, it has unequally important  packets  such that some packets contribute more to the video quality compared to other packets.  Second, it has a hard deadline such that the  packets need to be decoded on-time to be usable  at the applications.  The video  packets are  broadcasted from a  central station  to the devices over long-range wireless channels. However,  the  devices  receive  partial content in those transmissions due to erasures in  wireless  channels. To recover the missing packets, the devices communicate with each other  using their short-range wireless channels. Moreover, depending on the location of a device, it can be connected to all other devices directly (i.e., single-hop transmission) or via  intermediate devices (i.e., multi-hop transmissions).  Fig. \ref{fig:cc} shows an example of a heterogenous wireless network where devices   use  their cellular and short-range interfaces simultaneously.


Network coding has shown great potential to improve quality of services  for video streaming  applications in wireless networks \cite{yan2011weakly,firooz2013data,zhang2011mbms,wang2014minimizing,dong2013delay,zeng2012joint,tassi2014resource,thomos2014adaptive,vukobratovic2012unequal}.
In particular, random linear network coding (RLNC)    minimizes the number of transmissions required for wireless broadcast of  a set of packets \cite{tassi2014resource,thomos2014adaptive,vukobratovic2012unequal}. However, this throughput  benefit of RLNC comes at the expense of high decoding delay, high packet overhead, and high  encoding and decoding complexities. On the other hand, \emph{instantly decodable network coding} (IDNC) has drawn significant attention  due to its several attractive properties
\cite{li2011optimal,zhan2011coding,keller2008online,le2013instantly,sorour2012completion,li2011capacity,muhammad2013instantly}.
IDNC generates  coded packets that are immediately  decodable  at the devices. This instant decodability property allows a progressive improvement in the video quality as the devices decode more packets. Furthermore, the encoding process of IDNC is performed using simple XOR operations. This  reduces   packet overhead required for coefficient reporting. The decoding process of IDNC is also  performed using XOR operations, which is suitable for implementation in small  devices.

\ifCLASSOPTIONonecolumn
\begin{figure}[t]
        \centering
        \includegraphics[width=8cm,height=5cm]{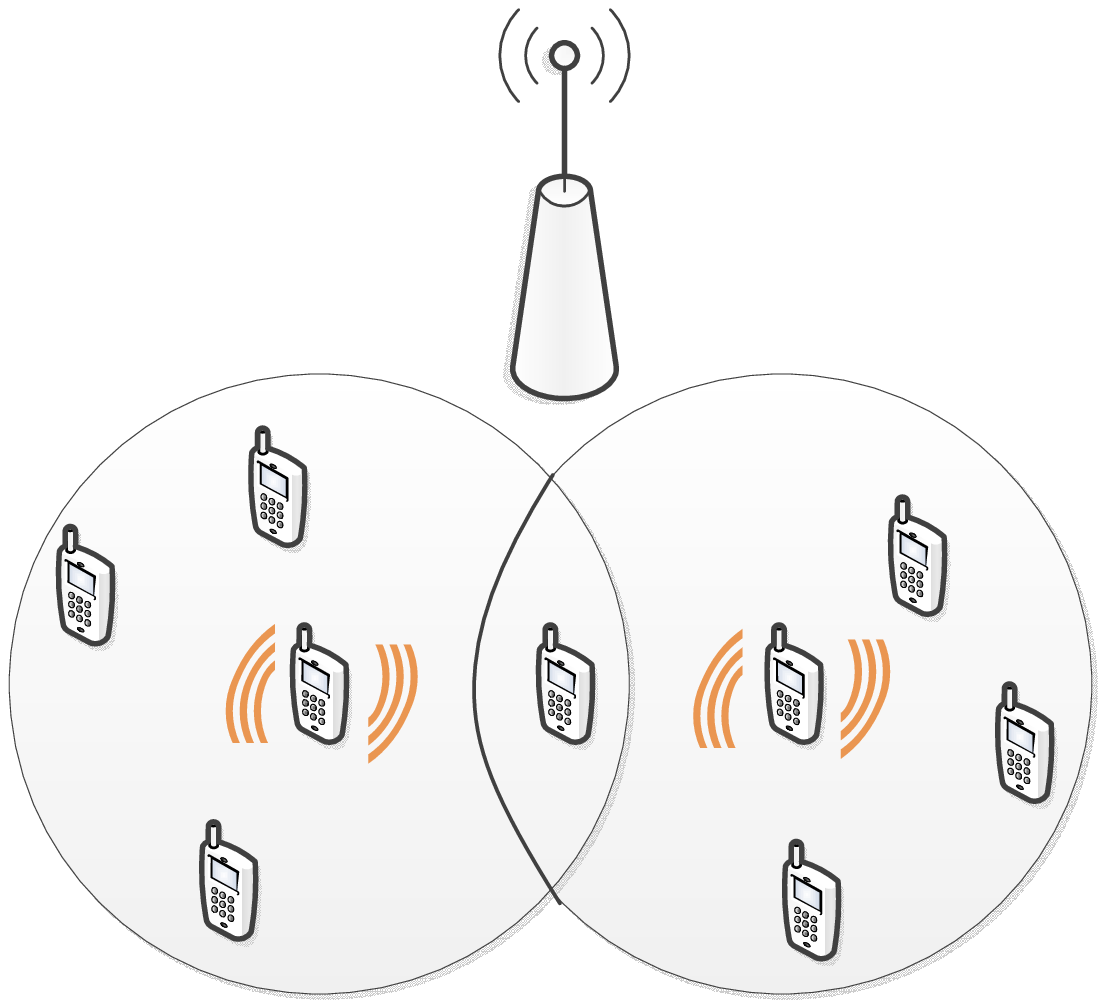}
        \caption{Devices   use  their cellular and short-range interfaces simultaneously.} \label{fig:cc}
\end{figure}
\else
\begin{figure}[t]
        \centering
        \includegraphics[width=\columnwidth]{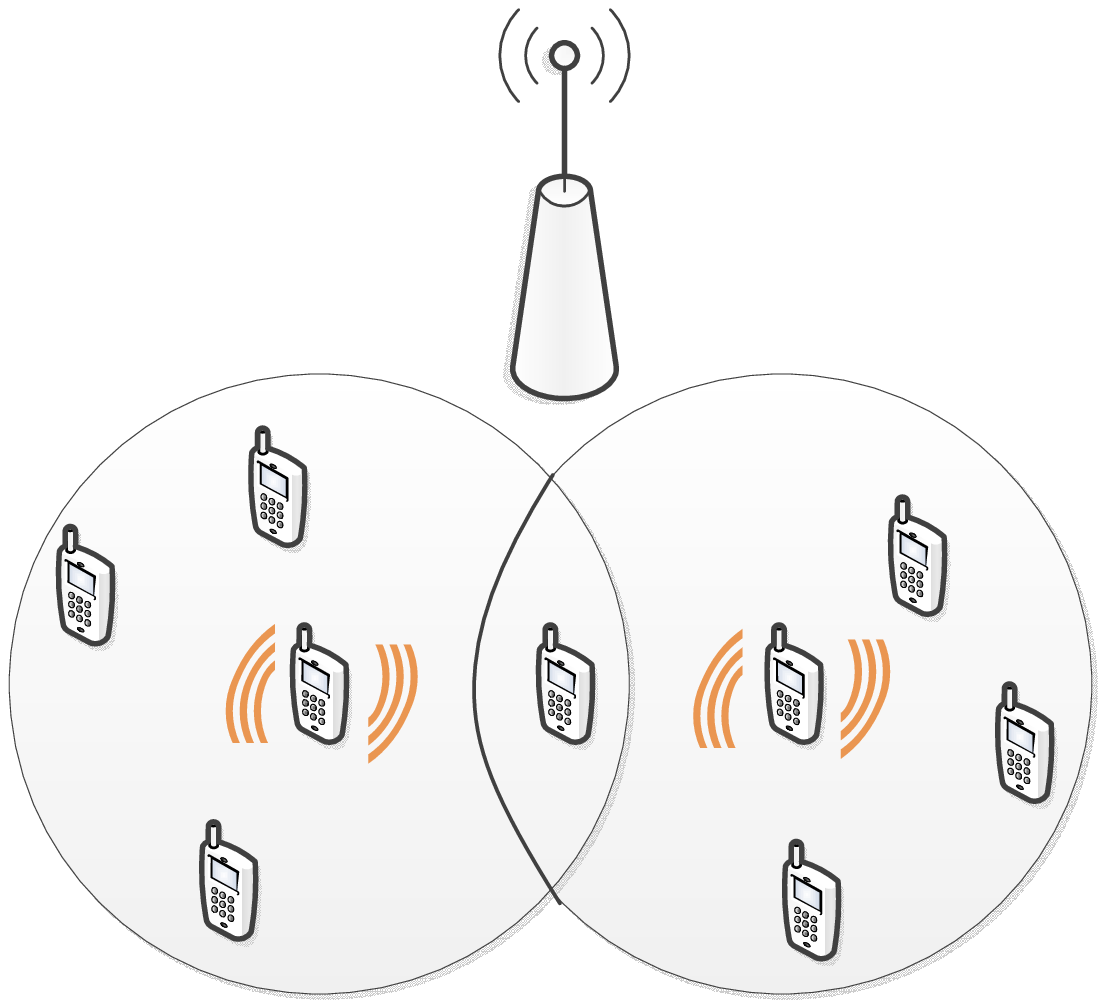}
        \caption{Devices   use  their cellular and short-range interfaces simultaneously.} \label{fig:cc}
\end{figure}
\fi
In this paper,  we are interested in designing an efficient  IDNC framework that minimizes the mean video distortion  before the deadline in a  partially connected device-to-device (D2D)  network.  In such  scenarios, IDNC framework needs to  take into account the unequal   importance  of video  packets,  hard deadline,  erasures of wireless channels, and coding and  transmission conflicts  in making decisions. In this context, our main contributions can be summarized as follows:
\begin{itemize}
\item   We introduce  a novel IDNC graph that represents both coding and transmission conflicts of a partially connected  D2D network with one common transmission channel. The representation of  transmission conflicts along with the well-known coding conflicts in one graph  were suggested in \cite{habob2015conflict,shnaiwer2015femto} for  distributed storage and femtocaching-assisted networks  for  transmissions over orthogonal channels. However, the representation of transmission and coding  conflicts in one  graph  for  a partially connected  D2D network with devices  all transmitting over one common  channel is not trivial and is novel to this paper. Indeed, this novel graph  representation has to  account for  the coverage zones of different devices, potential collisions  over the common channel,  each device  cannot  transmit and receive concurrently and the packet reception  at a device is subject to interference from simultaneous  transmissions of multiple devices.

\item Using the video  characteristics  and   the new  IDNC graph, we  formulate the problem of minimizing the mean video distortion  before the deadline as a finite horizon Markov decision process (MDP) problem. Our MDP formulation is a sequential decision making process in which the decision is made at the current time slot and takes into account   the coding opportunities at  the successor time slots so that the devices experience the minimum  video distortion  at the end of the deadline.
    The Markov decision process was also  used  in \cite{sorour2012completion,nguyen2007multimedia}  for  point to multi-point  networks, where the central station always transmits packets to the devices.  However, the MDP formulation  for  a partially connected  D2D network is different compared to those in \cite{sorour2012completion,nguyen2007multimedia} since it takes into account the fact that  a set of devices transmit XOR packet combinations  simultaneously and  another set of devices receive a single transmitted packet (i.e., free from transmission conflicts)  from  the  transmitting devices.

\item We further design a  two-stage  maximal independent set (TS-MIS) selection algorithm, which has much lower modelling and computational complexities compared to the MDP formulation. This is  a greedy  approach since it makes  decision  at  the current   time slot without going through all possible  future  situations before the deadline. However, this algorithm is designed following the properties of the minimum video distortion problem in a partially connected D2D network.

\item We use a real video sequence to  evaluate the performance of different  algorithms. Simulation results  show that our proposed IDNC algorithms improve the received video quality compared to  the IDNC algorithms in \cite{le2013instantly,keshtkarjahromi2015content,douik2014delay} that were not particularly designed for a real-time video sequence  and a  partially connected D2D  network.
\end{itemize}

The rest of this paper is organized as follows. We discuss the related works in Section \ref{literature} The system model is  described in Section \ref{tools}.   Section \ref{IDNCgraph} defines the novel IDNC graph.   We formulate the minimum video distortion problem into an MDP framework   in Section \ref{formulation} and design a TS-MIS selection algorithm  in Section~\ref{two-stage}. Section \ref{sec:distortion} describes the calculations for the importance of individual video packet.  Simulation results are presented   in  Section \ref{results}.  Finally,  Section \ref{conclusion} concludes the paper.

\section{Related Works} \label{literature}
In this section, we first discuss  the related network coding schemes designed for point to multi-point (PMP) networks (i.e., the central station  is responsible to transmit all packets to all devices) and then discuss the related network coding schemes designed   for fully connected D2D networks (i.e., each device is directly connected to all other devices) and partially connected D2D networks as considered in this paper.

\subsection{Point to Multi-Point (PMP)  Networks}
Numerous IDNC schemes    have been developed  to meet different requirements of  video streaming applications \cite{keller2008online,le2013instantly,sorour2012completion,li2011capacity,nguyen2007multimedia,seferoglu2009video,muhammad2013instantly}.   In particular, the authors in  \cite{keller2008online,le2013instantly} considered IDNC for  wireless broadcast of  a set of packets and  serviced the maximum number of devices with any new  packet in each time slot. Moreover,  the authors in \cite{sorour2012completion} addressed the problem of minimizing the number of time slots required for broadcasting a set of packets in IDNC systems  and formulated the problem into a stochastic shortest path (SSP) framework. However, the works in \cite{keller2008online,le2013instantly,sorour2012completion} neither considered explicit packet delivery deadline nor considered unequal importance of video packets.

Several other works including \cite{li2011capacity,nguyen2007multimedia,seferoglu2009video,muhammad2013instantly} considered video streaming  applications with unequally important packets. The work in \cite{li2011capacity}   proposed an IDNC scheme that is asymptotically throughput optimal for the three-device system subject to sequential packet delivery deadline constraints.  Moreover, the works in \cite{nguyen2007multimedia,seferoglu2009video}  determined the importance of each video packet based on its contribution to the video quality and proposed IDNC  schemes to  maximize the overall video quality at the devices.  The aforementioned works \cite{keller2008online,le2013instantly,sorour2012completion,li2011capacity,nguyen2007multimedia,seferoglu2009video,muhammad2013instantly} developed IDNC schemes for  conventional PMP networks, which are fundamentally different from  partially connected D2D networks considered in this paper.

\subsection{Fully Connected D2D Networks}

The  network coded D2D  communications   have drawn a significant attention over the past several years to take advantages of both network coding and devices' cooperation.   The works in \cite{el2010coding,sprintson2010randomized,milosavljevic2011deterministic} incorporated algebraic network coding  for D2D communications at the packet level.
In particular, the  authors   in  \cite{el2010coding}  provided upper and lower bounds on the number of time slots required for recovering all the missing packets at the devices. Furthermore, the authors in \cite{sprintson2010randomized} proposed a randomized algorithm that has a high probability of achieving the minimum number of time slots. However, the works in  \cite{el2010coding,sprintson2010randomized,milosavljevic2011deterministic}  neither   considered  erasure channels nor considered addressing  the hard deadline for high importance video packets.

Several other works including \cite{aboutorab2013instantly,karim2014decoding,keshtkarjahromi2015content} adopted  IDNC for  D2D communications.  In  \cite{aboutorab2013instantly,karim2014decoding}, the authors selected a transmitting device and its XOR packet combination  to service a large  number of  devices with any new packet in each time slot.  Moreover, the authors  in \cite{keshtkarjahromi2015content} prioritized packets based on their  contributions to the video quality as in \cite{nguyen2007multimedia,seferoglu2009video} and  proposed a joint device and packet selection algorithm that maximizes the overall  video quality after the current   time slot. The aforementioned  works \cite{el2010coding,sprintson2010randomized,milosavljevic2011deterministic,aboutorab2013instantly,karim2014decoding,keshtkarjahromi2015content}  developed network coding schemes for a  fully connected D2D  network.  This fully connected D2D network   is not always practical  due to the limited  transmission range of devices. Consequently,  in this paper,  we consider a partially connected D2D network, which is   more general and includes the fully connected D2D network as a special case. Unlike a single transmitting device in a fully connected D2D network, multiple devices  can transmit simultaneously in a partially connected D2D network without causing  transmission  conflicts.

\subsection{Partially Connected D2D Networks}
In the context of  partially connected  networks, the related works to our work   are \cite{courtade2014coded,gonen2012coded,douik2014delay,tajbakhsh2014instantly}. In particular, the authors  in \cite{courtade2014coded} provided various necessary and sufficient conditions that characterize the number of  transmissions required  to recover all missing packets at all devices.  The authors in   \cite{gonen2012coded} continued the work in \cite{courtade2014coded} and  showed that solving the   minimum  number of transmissions problem  exactly or even approximately is computationally intractable. Moreover, the  authors in \cite{courtade2014coded,gonen2012coded} adopted  algebraic network coding  in large finite fields. Unlike  the works in \cite{courtade2014coded,gonen2012coded}, we consider erasure  channels, XOR based network coding,  explicit packet delivery deadline and  unequal importance of video packets.

The works in \cite{tajbakhsh2014instantly,douik2014delay} adopted  IDNC  for a partially connected D2D network  and   addressed the problem of servicing  a large number of devices with any new packet in each   time slot. However, these works are not readily compatible with the real-time video sequence that has a hard deadline and unequally important video packets. In contrast to \cite{tajbakhsh2014instantly,douik2014delay}, we introduce a novel  IDNC graph that represents all feasible coding and transmission conflict-free decisions in one unified framework and develop an efficient IDNC framework  that prioritizes the distribution of   high importance video packets to all devices   before the deadline.

\section{System Model}\label{tools}
We consider  a wireless network with a   set of  $M$ devices $\mathcal{M} = \{R_1,...,R_M\}$.\footnotemark \footnotetext{Throughout this paper, we use calligraphic letters to denote sets and their corresponding capital letters to denote the cardinalities of these sets.} Each device in $\mathcal M$  is interested in receiving a set of $N$ source packets  $\mathcal{N} = \{P_1,...,P_N\}$.  Packets are transmitted in two phases. The first phase consists of the initial $N$ time slots, in which   a central station (e.g., a base station)  broadcasts the packets from  $\mathcal N$ in an uncoded manner.  However, a subset of  devices from $\mathcal{M}$  receive each broadcasted  packet due to erasures in long-range wireless channels. We assume that at least one device from $\mathcal{M}$ receives  each broadcasted  packet. 

\newtheorem{remarks}{\textbf{Remark}}

The  second phase starts after $N$  time slots (referred to as a \emph{D2D phase}), in which  the devices cooperate with each other to recover their missing packets using short-range wireless channels.   There is a limit on the  number of allowable  time slots $\Theta$  used in the D2D phase as the deadline for delivering $N$ packets  expires after $\Theta$ D2D  time slots. This deadline constraint arises from the minimum delivery delay requirement in real-time video streaming applications.  At any D2D  time slot $t \in [1, 2, ...,\Theta],$  we can compute  the number of remaining time slots for delivering $N$ packets  as, $Q = \Theta - t + 1.$  A device  can either transmit  or listen to a packet in each D2D time slot.



We consider a partially connected  network, where  a device is connected to another device  directly (i.e., single hop)  or via intermediate devices (i.e., multiple hops). The packet reception probabilities of all channels connecting  all pairs of devices  is stored in   an $M \times M$  \emph{symmetric connectivity matrix} (SCM) $\mathbf{Y} = [y_{i,k}], \forall (R_i, R_k) \in \mathcal M$, such that:
 \begin{equation}\label{eqn:Sta}
  y_{i,k} =
   \begin{cases}
    1 - \epsilon_{i,k}  & \text{if  $R_i$ is directly connected to  $R_k$},  \\
    0 & \text{otherwise}.
   \end{cases}
 \end{equation}
\begin{align}
y_{i,i} = 1 , \;\;\;\forall R_i \in \mathcal M.
\end{align}
Here, a  packet transmission  from device $R_i$ to device $R_k$ is subject to an independent Bernoulli  erasure with probability $\epsilon_{i,k}$.  We assume reciprocal  channels such as  $\epsilon_{i,k} = \epsilon_{k,i}$.  A channel connecting  a pair of devices is independent, but not necessarily identical,  to another channel connecting  another pair of devices. In fact, a device $R_i \in \mathcal M$ is  directly connected to a subset of devices in $\mathcal M$   depending on  the location of the  device in the network.

\newtheorem{definitions}{\textbf{Definition}}

\begin{figure}[t]
        \centering
\begin{tikzpicture}
\tikzstyle{vertex}=[auto=left,circle,fill=black!25,minimum size=30pt,inner sep=0pt]

\node[vertex,minimum size=25pt,color=green!25,text=black] (n9)   {$R_1$};
\node[vertex,minimum size=25pt,color=green!25,text=black] (n10) [right = 1.8cm] at (n9) {$R_2$};
\node[vertex,minimum size=25pt,color=green!25,text=black] (n11) [right = 1.8cm] at (n10) {$R_3$};
\node[vertex,minimum size=25pt,color=green!25,text=black] (n12) [right = 1.8cm] at (n11) {$R_4$};
\foreach \from/\to in {n9/n10,  n10/n11, n11/n12}
\draw [line width=1pt] (\from) -- (\to);
\node (S2)[right = .95cm , above = .5cm]  at (n9) {$0.84$};
\node (S3)[right = .95cm , above = .5cm]  at (n10) {$0.75$};
\node (S4)[right = .95cm , above = .5cm] at (n11) {$0.91$};
\end{tikzpicture}
\caption{A line network corresponding to SCM in \eqref{eqn:localIDNC}.}\label{fig:line}
\end{figure}
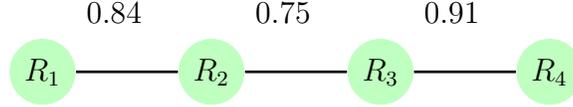

\newtheorem{examples}{\textbf{Example}}
\begin{examples}
An example of  SCM  with $M = 4$ devices is given as follows:
\begin{equation}\label{eqn:localIDNC}
\mathbf{Y} = \begin{pmatrix}
   1 & 0.84 & 0 & 0\\
   0.84 & 1 & 0.75 & 0\\
   0 & 0.75 & 1 & 0.91\\
   0 & 0 & 0.91 & 1\\
\end{pmatrix}.
\end{equation}
The SCM in \eqref{eqn:localIDNC} represents  a line network  shown in Fig. \ref{fig:line}. In this example, device $R_1$ is  not directly connected to  device $R_3$  and thus, $y_{1,3} = 0$. Moreover, device $R_1$ is directly connected to device $R_2$    with packet reception  probability $y_{1,2} = 1-\epsilon_{1,2} = 0.84$.
\end{examples}

\begin{definitions}
(Coverage Zone) The coverage zone of transmitting device $R_i$ (denoted by $\mathcal{Y}_i$) is defined as the set of neighboring  devices that are   directly connected  to it using  short-range wireless channels. In other words,  $ \mathcal Y_i =\{R_k \mid y_{i,k} \neq 0 \}$.
\end{definitions}

\begin{definitions}
(Transmission Conflict) A transmission conflict is experienced by  a device when  it  belongs to the  coverage zones  of multiple transmitting devices.  In other words, when  two neighboring devices $R_i$ and $R_r$ of   device $R_k$  transmit simultaneously, their transmissions will collide and  device $R_k$ will not be able to receive any of these transmissions successfully.
\end{definitions}

After each  time slot, the   reception  status of all packets  at all devices is stored   in  an $M\times N $  \emph{global status matrix (GSM)} $\mathbf{F} = [f_{k,l}],$ $\; \forall R_k\in\mathcal{M}, P_l\in \mathcal{N}$, such that:
 \begin{equation}
  f_{k,l} =
   \begin{cases}
    0 & \text{if packet $P_l$ is received by device $R_k$},  \\
    1 & \text{if packet $P_l$ is missing at device $R_k$.}
   \end{cases}
 \end{equation}

\begin{examples}
An example of GSM  with $M = 4$ devices and $N = 3$ packets is given as follows:
\begin{equation}\label{eqn:globalIDNC}
\mathbf{F} = \begin{pmatrix}
   1 & 1 & 0\\
   0 & 1 & 1\\
   0 & 0 & 1\\
   1 & 0 & 1\\
\end{pmatrix}.
\end{equation}
\end{examples}
According to  the GSM $\mathbf{F}$, the following  two sets of packets can be attributed to each device $R_k \in \mathcal{M}$ at any given   time slot $t$:
\begin{enumerate}
\item The \emph{Has set} ($\mathcal{H}_k)$ of device $R_k$ is defined as the set of packets that are successfully  received by device $R_k$. In \eqref{eqn:globalIDNC}, the Has set of device $R_1$ is $\mathcal H_1 = \{P_3\}$.
\item The \emph{Wants set} ($\mathcal{W}_k)$ of  device $R_k$ is defined as the set of  packets that are missing at device $R_k$. In other words, $\mathcal{W}_k = \mathcal N \setminus \mathcal{H}_k $. In \eqref{eqn:globalIDNC}, the Wants set of device $R_1$ is $\mathcal W_1 = \{P_1, P_2\}$.
\end{enumerate}

The cardinalities of $\mathcal{H}_k$ and   $\mathcal{W}_k$   are denoted  by $H_k$ and  $W_k$, respectively.  The set of devices having \emph{non-empty Wants sets} is denoted by $\mathcal M_w$. This set can be defined as:  $\mathcal M_w =\{R_k\mid \mathcal{W}_k \neq \varnothing\}$.  At any given time slot $t$, a device $R_k$ in $\mathcal{M}_w$  belongs to one of the following  two sets:
\begin{itemize}
\item The \emph{critical set} of devices ($\mathcal{C}$) is  defined as   the set of devices with the number of  missing packets     being greater than or equal to the number of remaining  $Q$ time slots (i.e., $W_k \geq Q, \forall R_k \in    \mathcal{C}$).

\item The \emph{non-critical set} of devices ($\mathcal{A}$)     is defined as the set of devices with the number of  missing packets  being less than the number of remaining $Q$  time slots (i.e., $W_k < Q, \forall     R_k \in \mathcal{A}$).
\end{itemize}
In fact, $\mathcal{C}(t) \cup \mathcal{A}(t) = \mathcal M_w(t)$.

\begin{definitions}
(Instantly Decodable Packet) A transmitted packet is instantly decodable for device $R_k$ if it contains exactly one source packet from $\mathcal W_k$.
\end{definitions}
\begin{definitions}\label{de:targ}
(Targeted Device) Device $R_k$  is targeted by transmitting device $R_i$ with packet $P_l$ at   time slot $t$ when  device $R_k$ belongs to the coverage zone of a single transmitting   device $R_i$   and  will immediately decode  packet $P_l$  upon  receiving  the transmitted  packet from device $R_i$.
\end{definitions}

\begin{definitions}
(Individual Completion Time) At any time slot $t$,  individual completion  time  of  device $R_k$  (denoted by $T_{W_k}$) is  the total  number of time slots required to decode all the missing packets in   $W_k$.
\end{definitions}
Individual completion time of device $R_k$ for $W_k$ missing packets can be $T_{W_k} = W_k, W_{k}+1,...$ depending on the number of time slots in which this device is targeted with a new packet (i.e., satisfies Definition \ref{de:targ}) and the channel erasures experienced by this device in those transmissions.
\begin{definitions}
(Individual Completion Times of All Non-critical Devices) At any time slot $t$, individual completion times of all non-critical devices  (denoted by $T_A$) is the total number of time slots  required to deliver all the missing packets to all non-critical devices in $\mathcal A$.
\end{definitions}

\begin{definitions}
(Transmission Schedule) A transmission schedule $\mathcal L = \{\kappa(t)\}, \forall t \in \{1,...,\Theta\}$ is defined as the set of transmitting devices and  packet combinations at every time slot $t$ before the deadline. Furthermore,  $\mathbf{L}$ is the set of all possible transmission schedules and $\mathcal L \in \mathbf{L}$.
\end{definitions}

\subsection{Centralized Protocol for Implementing the System }

As a potential protocol, we now discuss the possible  implementation processes of the  IDNC system in a centralized fashion.\footnotemark \footnotetext{A distributed approach
can be adopted to make a decision at each device separately. Many works on distributed approaches were referred in  \cite{al2013energy,tajbakhsh2014instantly}.} In this case, the central  station  forms the SCM  $\mathbf Y$ and the  GSM $\mathbf F$, and  coordinates the global decision making process  in each time slot.

\subsubsection{Coverage Zone}

The  devices exchange  Hello messages among themselves in order to  determine their   coverage zones (i.e., neighbouring  devices).  Each device broadcasts one bit Hello message. Other $O(M-1)$ neighboring  devices generate  one bit response message. Consequently,  a device discovers its coverage zone using $M$ bits. The coverage zones of all $M$ devices in the network can be discovered using $M^2$ bits. Since the locations of all devices in the network  are static with respect to the delivery deadline of the video sequence, the communication overhead of $M^2$ bits is   required only once.

\subsubsection{Packet Reception Probability}
In this paper, the network coding  is performed at the network layer. With an efficient channel coding  performed at the physical layer,   an abstraction of channel model at the network layer is often considered, where  a transmitted  packet is either  received or  lost with an  average  erasure  probability. This channel  erasure probability  is a slowly changing parameter in the network and can be estimated  based on the  test (or the past)   packet reception performance over the channel. Once  the packet reception probabilities connecting a device to other devices are estimated, the device   sends this information  to the central  station. A channel erasure probability  can be represented  using $\lceil \log_2 100  \rceil$ bits, where 100 is  the maximum erasure probability in percentage. Since each of $M$ devices sends $M-1$ channels' information connecting this device to other $M-1$ devices, the overall communication overhead     is $M^2\lceil \log_2 100  \rceil$ bits. Using this information, the central station  forms the SCM  $\mathbf Y$.

\subsubsection{GSM Update}
Each device  sends a positive/negative acknowledgement to the central station  indicating a received/lost packet. Note that a device needs to use one bit to acknowledge  a received packet. Since there are $M$ devices in the network, the overall communication overhead from feedback is $M$ bits per time slot. With the feedback reception, the central station updates the GSM $\mathbf F$ in each time slot.

\subsubsection{Centralized Decision}
In each time slot, the central station selects  a set of  transmitting devices and their packet combinations using an IDNC algorithm. It then informs the  transmitting devices  separately about the   packet combinations and uses   the indices of  individual  packets. In fact, a packet combination  can be formed XORing   $O(N)$ individual  packets. The central station  sends  a bitmap of  $N$  bits to  each transmitting device, where the entries with 1's are the indices of the source  packets that are  XORed together.  In a partially connected  D2D network, there can be $O(\frac{M}{2})$ transmitting devices since a device cannot receive and transmit simultaneously. The overall communication overhead to inform $O(\frac{M}{2})$ transmitting devices about their packet combinations is $\frac{MN}{2}$ bits, which is negligible compared to  the typical  size of a packet in wireless   networks.



\subsection{Importance of Individual Packet} \label{sec:packetI}
The importance of individual  packet in a video sequence  can be determined by the source and can be marked on a special field of the packet header. This field can be part of the  real-time transport protocol (RTP)   header or  the network coding header   \cite{seferoglu2009video}.
To compute the importance of packet $P_l$, we follow a  similar approach as in \cite{seferoglu2009video,nguyen2007multimedia} and decode the entire video sequence with this packet missing and assign the resulting distortion to the importance value  of this packet. This is an approximation as the actual distortion of a packet depends on the reception status of prior and subsequent packets  at the devices. Having defined  the importance  of individual packets,  we calculate the individual   video distortion of device $R_k$ at  time slot $t$   as:
\begin{equation} \label{eq:sumD}
d_{k}^{(t)} = \sum_{P_l \in \mathcal W_k} \delta_{k,l}
\end{equation}
where $\delta_{k,l}$ is the importance of missing packet $P_l$ at device $R_k$.
Here, we consider that distortions caused by the loss of multiple packets at a device are additive, which is accurate   for sparse losses. Nonetheless,  these  approximations   allow us to separate the total distortion of a video sequence  into a set of  distortions corresponding to individual  packets and optimize the   decisions for individual packets. To compute the received video quality at the devices,  we capture  the  correlations of the packets in a video sequence. We use these correlations to compute the actual video distortion at a device  resulting from its missing packets  at the end of  the deadline. These practical aspects in computing the received video quality at the devices   will be further  explained   in Section \ref{sec:distortion}.


\section{Novel IDNC Graph}\label{IDNCgraph}

In this section, we  define a novel IDNC graph $\mathcal G(\mathcal V, \mathcal E)$ to represent both coding and transmission conflicts in one unified framework and  select a set of transmitting devices and their  XOR packet combinations in each D2D time slot. A transmission conflict occurs  due to the simultaneous   transmissions from multiple devices to a  device in their coverage zones. Moreover, a coding conflict occurs  due to the instant decodability constraint.

\subsection{Vertex Set}

To define vertex set $\mathcal V$ of  IDNC graph  $\mathcal G$,  given GSM $\mathbf F$ at  time slot $t$,  we  form an $Y_i \times H_i$  \emph{local status matrix (LSM)} $\mathbf{F}_{i} = [f_{k,l}],$ $\; \forall R_k\in\mathcal{Y}_i, P_l\in \mathcal{H}_{i}$,  for  a device $R_i \in \mathcal M$ such that\footnotemark \footnotetext{The number of devices in the coverage zone of  device $R_i$ is $Y_i = |\mathcal{Y}_i|$.}:
  \begin{equation}
  f_{k,l} =
   \begin{cases}
    0 & \text{if packet $P_l$ is received by device $R_k$},  \\
    1 & \text{if packet $P_l$ is missing at device $R_k$.}
   \end{cases}
 \end{equation}
Note that the rows in LSM $\mathbf{F}_{i} $ represent the  devices which  are in the coverage zone of  device $R_i$ and the columns in LSM $\mathbf{F}_{i} $ represent   the packets  in the Has set of device $R_i$ which are used  for forming a transmitted  packet from device $R_i$. Fig. \ref{fig:LSM} shows four  LSMs  for  four devices corresponding to SCM in \eqref{eqn:localIDNC} and  GSM   in \eqref{eqn:globalIDNC}.

\ifCLASSOPTIONonecolumn

\begin{figure}
\centering
\tikzset{
    table nodes/.style={
        rectangle,
        draw=black,
        align=center,
        minimum height=7mm,
        text depth=0.5ex,
        text height=2ex,
        inner xsep=0pt,
        outer sep=0pt
    },
    table/.style={
        matrix of nodes,
        row sep=-\pgflinewidth,
        column sep=-\pgflinewidth,
        nodes={
            table nodes
        },
        execute at empty cell={\node[draw=none]{};}
    }
}

\begin{tikzpicture}

\matrix (first) [table,text width=7mm,name=table]
{ 1  \\
 0  \\
 0  \\
 };

\begin{scope}[xshift=-3.5 cm]
\matrix (second) [table,text width=7mm,name=table] {
 0  \\
  1  \\
 };
\end{scope}

\begin{scope}[xshift=3cm]
\matrix (second) [table,text width=7mm,name=table] {
 0 & 1  \\
 0 & 0 \\
 1 & 0 \\
};
\end{scope}

\begin{scope}[xshift=7cm]
\matrix (second) [table,text width=7mm,name=table] {
 0  \\
 0  \\
};
\end{scope}
%
\node   [below = 1.4cm]   (T_a){};

\node  (A_a) [right = -3.5cm, below = 0.1cm] at  (T_a){$\mathbf{F}_1$};
\node  (A_b) [right = 3.2cm] at  (A_a){$\mathbf{F}_2$};
\node  (A_c) [right = 2.5cm] at  (A_b){$\mathbf{F}_3$};
\node  (A_d) [right = 3.7cm] at  (A_c){$\mathbf{F}_4$};

\end{tikzpicture}
\caption{Four LSMs for four devices corresponding  to SCM in \eqref{eqn:localIDNC} and GSM in \eqref{eqn:globalIDNC}} \label{fig:LSM}
\end{figure}

\else

  \begin{figure}
\centering
\tikzset{
    table nodes/.style={
        rectangle,
        draw=black,
        align=center,
        minimum height=7mm,
        text depth=0.5ex,
        text height=2ex,
        inner xsep=0pt,
        outer sep=0pt
    },
    table/.style={
        matrix of nodes,
        row sep=-\pgflinewidth,
        column sep=-\pgflinewidth,
        nodes={
            table nodes
        },
        execute at empty cell={\node[draw=none]{};}
    }
}

\begin{tikzpicture}

\matrix (first) [table,text width=7mm,name=table]
{ 1  \\
 0  \\
 0  \\
 };

\begin{scope}[xshift=-1.5 cm]
\matrix (second) [table,text width=7mm,name=table] {
 0  \\
  1  \\
 };
\end{scope}

\begin{scope}[xshift=2cm]
\matrix (second) [table,text width=7mm,name=table] {
 0 & 1  \\
 0 & 0 \\
 1 & 0 \\
};
\end{scope}

\begin{scope}[xshift=4cm]
\matrix (second) [table,text width=7mm,name=table] {
 0  \\
 0  \\
};
\end{scope}
%
\node   [below = 1.4cm]   (T_a){};

\node  (A_a) [right = -1.5cm, below = 0.1cm] at  (T_a){$\mathbf{F}_1$};
\node  (A_b) [right = 1.2cm] at  (A_a){$\mathbf{F}_2$};
\node  (A_c) [right = 1.5cm] at  (A_b){$\mathbf{F}_3$};
\node  (A_d) [right = 1.7cm] at  (A_c){$\mathbf{F}_4$};

\end{tikzpicture}
\caption{Four LSMs for four devices corresponding  to SCM in \eqref{eqn:localIDNC} and GSM in \eqref{eqn:globalIDNC}} \label{fig:LSM}
\end{figure}

\fi

We generate a vertex for a  missing packet   in each  LSM  at  IDNC graph $\mathcal G$. In fact, for each  LSM $\mathbf F_{i}, \forall R_i \in \mathcal M$,  a vertex $v_{i,kl}$  is generated for a packet $P_l \in \{\mathcal{H}_i \cap \mathcal{W}_k\}, \forall R_k \in \mathcal Y_i$.\footnotemark \footnotetext{Note that vertex $v_{i, kl}$ represents a transmission from device $R_i \in \mathcal M$ to a neighboring device $R_k \in \mathcal Y_i$ with packet $P_l$.} In other words, a vertex is  generated for  a missing packet of another device in  $\mathcal Y_i$, which also belongs to the Has set $\mathcal{H}_i$  of potential transmitting device $R_i$.  Note that a missing  packet at a device can generate  more than one vertex  in  graph $\mathcal G$ since that packet can be present in multiple  LSMs. Once the vertices are generated in  IDNC graph $\mathcal{G}$, two vertices $v_{i, kl}$ and $v_{r, mn}$  are adjacent (i.e., connected) by  an edge due to  either a coding conflict  or a transmission conflict.


\subsection{Coding Conflicts}
Two vertices $v_{i, kl}$ and $v_{r, mn}$   are  adjacent by an edge
due to  a coding conflict  if one of the following two conditions holds:
\begin{itemize}
\item \textbf{C1}: $P_l \neq P_n$ and $R_k = R_m$. In other words,  two vertices are induced by different  missing packets $P_l$ and $P_n$ at the same  device $R_k$.
\item \textbf{C2}: $R_k \neq R_m$ and $P_l \neq P_n$ but $P_l \notin  \mathcal H_m$ or $P_n \notin  \mathcal H_k$. In other words, two different devices $R_k$ and $R_m$ require two different packets $P_l$ and $P_n$, but at least one of these two devices does not possess the other missing packet. As a result, that device cannot decode a new  packet from an XOR combination of $P_l \oplus P_n$.
\end{itemize}


\subsection{Transmission Conflicts}
Two vertices $v_{i, kl}$ and $v_{r, mn}$  are adjacent by an edge  due to  a transmission conflict  if one of  the following three  conditions holds:
\begin{itemize}
 \item \textbf{C3}: $R_i \neq R_r$ and $R_k = R_m \in \{\mathcal Y_{i} \cap \mathcal Y_{r}\}$. In other words,  two vertices representing the transmissions from two different devices $R_i$ and $R_r$  to the same device $R_k$ in  the coverage zones of both  transmitting devices $R_i$ and $R_r$.  This prohibits transmissions from two different  devices to  the same device  in the common coverage zone and  prevents  interference at that device from multiple transmissions.

 \item \textbf{C4}: $R_i \neq R_r$  and $R_k \neq R_m$ but $R_k \in \{\mathcal Y_{i} \cap \mathcal Y_{r}\} $ or $R_m \in \{\mathcal Y_{i} \cap \mathcal Y_{r}\}$. In other words,  two vertices representing the transmissions from two different devices $R_i$ and $R_r$  to  two   different  devices  $R_k$ and $R_m$, but  at least one of these two   devices $R_k$ and $R_m$ is in  the coverage zones of both  transmitting devices $R_i$ and $R_r$. This prohibits transmission from device $R_r$ to  device $R_m$ in the case of transmission from  device $R_i$ to device $R_k$, and vice versa.

 \item \textbf{C5}: $R_i \neq R_r$  but $R_i = R_m$ or $R_r = R_k$. In other words,  two vertices representing the transmissions from two different devices $R_i$ and $R_r$, but at least one of these two devices $R_i$ and $R_r$ is targeted by the other device. This prohibits  transmission from a device in the case of  that device is already targeted by another device, and vice versa. In other words, a device cannot be a transmitting device and a targeted device   simultaneously.
 \end{itemize}

\subsection{Maximal Independent Sets}

With this graph representation, we can define  all feasible  coding and transmission conflict-free decisions  by the set of all maximal independent sets in  IDNC graph $\mathcal G$.
\begin{definitions} \label{inde}
(Independent Set) An independent set or a stable set in a  graph is a set of pairwise non-adjacent vertices.
\end{definitions}
\begin{definitions}
(Maximal  Independent Set) A maximal  independent set  (denoted by $\kappa$) is an independent  set that cannot be extended by including one more vertex without violating pairwise non-adjacent vertex constraint. In other words, a maximal independent set  is an independent set that is not subset of any larger independent set \cite{west2001introduction}.
\end{definitions}

Each device can have at most one vertex in a maximal independent set $\kappa$ representing either  a transmitting device or  a targeted device. Moreover, the selection of a maximal independent set  $\kappa$ is equivalent to the selection of a set of transmitting devices $\mathcal Z (\kappa) = \{R_i | v_{i,kl} \in \kappa\}$ and  a set of targeted devices  $\mathcal X (\kappa) = \{R_k | v_{i,kl} \in \kappa\}$. Each of the selected transmitting devices   forms a  coded packet  by XORing the source packets identified by the vertices in $\kappa$ representing transmission from  that device.

\begin{figure}[t]
        \centering
\begin{tikzpicture}
\tikzstyle{vertex}=[auto=left,circle,fill=black!25,minimum size=30pt,inner sep=0pt]

\node[vertex,minimum size=25pt,color=green!25,text=black] (n9) [right = 8.2cm]  {$v_{1,23}$};
\node[vertex,minimum size=25pt,color=green!25,text=black] (n10) [right = 1.5cm] at (n9) {$v_{3,22}$};
\node[vertex,minimum size=25pt,color=green!25,text=black] (n11) [below = 1.3cm] at (n9) {$v_{2,11}$};
\node[vertex,minimum size=25pt,color=green!25,text=black] (n12) [below = 1.3cm] at (n10) {$v_{3,41}$};
\foreach \from/\to in {n9/n11,  n10/n11, n9/n10, n9/n12}
\draw [line width=1pt] (\from) -- (\to);
\node (S2)[right = 2.8cm , below = 0cm]  at (n10) {$\kappa_1=\{ v_{1,23}\}$};
\node (S3)[right = 2.8cm, below = 0.5cm]  at (n10) {$\kappa_2=\{ v_{2,11}, v_{3,41}\}$};
\node (S4)[right = 2.8cm, below = 1cm]  at (n10) {$\kappa_3=\{ v_{3,41}, v_{3,22}\}$};

\end{tikzpicture}
\caption{IDNC graph corresponding to   SCM in \eqref{eqn:localIDNC} and GSM in \eqref{eqn:globalIDNC}.}\label{fig:localIDNCgraph}
\end{figure}
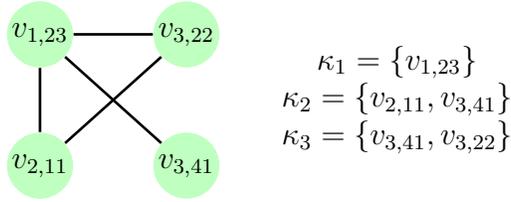

\begin{examples}
The new  IDNC graph $\mathcal G$ corresponding to  SCM in \eqref{eqn:localIDNC} and GSM in \eqref{eqn:globalIDNC}  is shown in Fig. \ref{fig:localIDNCgraph}.  The maximal independent sets of this graph are also listed in this figure.
\end{examples}

\section{Minimum  Video Distortion Problem Formulation}\label{formulation}

In this section, we first define   the minimum mean video distortion problem and  then formulate the problem into a finite horizon  Markov decision process (MDP) framework.

\subsection{Problem Description}
We now discuss the  characteristics of  the minimum video distortion   problem  and infer that  it  is a sequential decision making problem. In such a problem, the decision is made  at the current time slot and needs to  take into account all possible   GSMs and their  coding opportunities   at the successor time slots before the deadline.
First,  some packets  are needed  to be   exchanged via    multiple hops before the deadline due to the partial connectivity in the network. Therefore,  the decision at the current time slot needs to consider  that some devices  are able to quickly relay their received packets to a large number of other  devices  in the successor time slots due to having  large  coverage zones. Second, it is not always possible to target all the devices with a new packet due to the instant decodability constraint.  Moreover, servicing the largest number of devices with a new packet in the current  time slot may reduce the coding opportunities at the successor time slots, and results in  delivering a small number of  packets to the devices before the deadline.  Therefore, the  decision at the current time slot needs to  take into account the  coding opportunities   at the successor time slots before the deadline. Finally, the hard deadline constraint may limit  the number of delivered packets to the devices. Therefore, the decision maker needs to be adaptive to the deadline so that the received video packets before the deadline contribute to the maximum  video quality at the devices.

Based on all aforementioned  aspects, we can infer that  our problem is a sequential  decision making problem that not necessarily  minimizes the mean video distortion after the current time slot, but rather it   achieves the minimum mean video distortion  at the end of the deadline. Moreover,  due to the random nature of channel erasures,  our  system is a stochastic  system, in which there are many possible outcomes resulting from  a chosen  maximal independent set at the  current time slot. To define the minimum video distortion  problem, let us consider $d_k(\mathcal L)$ and $\mathcal H_k(\mathcal L)$ are the  individual  video distortion and the Has set  of device $R_k$  at the end of the deadline for a given  transmission schedule $\mathcal L$. Moreover,  $d_k^{(0)}$ is the initial  individual video distortion of device $R_k$ before starting   the D2D phase and can be computed following  \eqref{eq:sumD}.
With these results, we  define  the problem of  minimizing  the mean video distortion  at the end of  the deadline as  a transmission schedule selection problem such that:

\begin{align}\label{eq:minmax}
\mathcal L^* &= \arg\min_{\mathcal L \in \mathbf{L}}\left\{ \frac{\sum_{R_k \in \mathcal M} d_k(\mathcal L)}{M}\right\} \nonumber \\
&= \arg\min_{\mathcal L \in \mathbf{L}}\left\{ \sum_{R_k \in \mathcal M} d_k(\mathcal L)\right\} \nonumber \\
&= \arg\min_{\mathcal L \in \mathbf{L}}\left\{ \sum_{R_k \in \mathcal M}\left (d_k^{(0)} -  \sum_{P_l \in \mathcal H_k(\mathcal L)} \delta_{k,l}\right)  \right\} \nonumber \\
&= \arg\max_{\mathcal L \in \mathbf{L}}\left\{ \sum_{R_k \in \mathcal M}   \sum_{P_l \in \mathcal H_k(\mathcal L)} \delta_{k,l}  \right\}.
\end{align}
The optimization problem  in \eqref{eq:minmax} can be formulated   using  a finite horizon Markov decision process  and the optimal transmission schedule can be found using the backward induction algorithm, which will be shown in the following two subsections.

\subsection{MDP Formulation}
We  formulate the problem of minimizing the mean video distortion before  the deadline as a finite horizon Markov decisions process (MDP) problem, which  models our decision based stochastic dynamic systems with a finite number of steps.

\begin{enumerate}
\item \emph{Horizon}: The  number of  time slots $\Theta$ used in the D2D phase, over which the decisions are made. The MDP problem is a finite horizon problem with $\Theta$ time slots.

\item \emph{State Space} $\mathcal{S}$: States are defined by all possibilities of GSM $\mathbf{F}$ that may occur during the D2D  phase.  GSM corresponding to state  $ s \in \mathcal{S}$ is represented by $\mathbf{F}(s)$. We can characterize each state $s$  according to its Has and  Wants vectors, $\mathbf{h}(s) = [H_1(s),...,H_M(s)]$ and $\mathbf{w}(s) = [W_1(s),...,W_M(s)]$.  The state at the  starting of the D2D phase is denoted by  $s_a$ and its Has and Wants vectors are denoted by  $\mathbf{h}(s_a) = [H_1(s_a),...,H_M(s_a)]$ and $\mathbf{w}(s_a) = [W_1(s_a),...,W_M(s_a)]$.

    Given GSM $\mathbf{F}$ is an $M\times N$ binary matrix,  the size of the state space is $|\mathcal{S}| = O(2^{MN})$. However, the devices  receive a subset of packets from $\mathcal N$ in the initial $N$ time slots from  the central station. We can conclude that the size of the state space for D2D phase is $|\mathcal{S}| = 2^{MN} - 2^{(\Sigma_{R_i \in \mathcal M}H_i(s_a))}$.

\item \emph{Action Space }$\mathcal{A}(s)$: The action space   for each state $s$  consists of the set of  all possible maximal independent sets in   IDNC graph $\mathcal G(s)$. The size of the action space for a given   state $\mathbf F(s)$ is $|\mathcal A(s)| = O(3^{|\mathcal V|/3})$ \cite{west2001introduction}, where $|\mathcal V|$ is the size of the vertex set $\mathcal V$ in graph $\mathcal G(s)$.

\item \emph{State-Action Transition Probability $\mathcal{P}_{a}(s,\acute{s})$}: The state-action transition probability $\mathcal{P}_{a}(s,\acute{s})$ for an action $a = \kappa(s)$  can be defined based on the possibilities of the variations in GSM $\mathbf{F}(s)$ from state $s$ to the successor  state $\acute{s}$. With  action $\kappa(s)$, the system  transits to the successor  state $\acute{s}$ depending on the targeted  devices in $\kappa(s)$  and the packet reception probabilities of  the targeted devices. In other words, successor state $s' \in \mathcal S(s,a)$ such that $\mathcal S(s,a) = \{\acute{s}|\mathcal{P}_a(s,\acute{s}) > 0\}$. To define   $\mathcal{P}_{a}(s,\acute{s})$, we first introduce the following two sets:

\begin{equation}
\mathcal{T} = \{ R_k | R_k \in \mathcal{X}(\kappa),  W_k(\acute{s})  =  W_k(s)-1 \}
\end{equation}
\begin{equation}
\tilde{\mathcal{T}} = \{R_k | R_k  \in \mathcal{X}(\kappa), W_k(\acute{s})  =  W_k(s)\}
\end{equation}
Here, the first set $\mathcal{T}$ includes the targeted  devices whose Wants sets have  decreased from state $s$ to the successor state $\acute{s}$ due to  successful packet receptions.  The second set $\tilde{\mathcal{T}}$ includes the targeted devices  whose Wants sets have  remained unchanged  due to packet  losses. Using  these two  sets and considering all transmissions are independent of each other, we can express   $ \mathcal{P}_{a}(s,\acute{s})$  as follows:
\ifCLASSOPTIONonecolumn
\begin{equation}
  \mathcal{P}_{a}(s,\acute{s})  = \prod_{R_k \in \mathcal{T}: v_{i,kl} \in \kappa(s)} (1 - \epsilon_{i,k}) \times \prod_{R_k \in \tilde{\mathcal{T}}: v_{i,kl} \in \kappa(s)}  (\epsilon_{i,k})
\end{equation}
\else
\begin{align}
 & \mathcal{P}_{a}(s,\acute{s}) \nonumber \\
  & = \prod_{R_k \in \mathcal{T}: v_{i,kl} \in \kappa(s)} (1 - \epsilon_{i,k}) \times \prod_{R_k \in \tilde{\mathcal{T}}: v_{i,kl} \in \kappa(s)}  (\epsilon_{i,k})
\end{align}
\fi

\item \emph{State-Action Reward}: Having required the minimum  mean video distortion at the end of the deadline, at state $s$, the expected  reward $\bar{r}_k(s,a)$ of action $a = \kappa(s)$ on  each device $R_k \in \mathcal{M}_w(s)$ is defined as the  expected video  distortion reduction  at device $R_k$ at the successor state $s'$.  We can calculate   the expected reward of action $a = \kappa(s)$ on  each targeted  device $R_k \in \mathcal{X}(a)$  as $\bar{r}_k(s,a|v_{i,kl} \in \kappa(s)) =    \delta_{k,l}(1-\epsilon_{i,k})$.  On the other hand, we can   define   the expected reward of action $a = \kappa(s)$ on  each ignored  device $R_k \in \{\mathcal M_w(s) \setminus \mathcal X(a)\}$  as $\bar{r}_k(s,a|R_k \in \mathcal M_w(s) \setminus \mathcal X(a)) =   0$. With these results, the total expected  reward of action $a \in \mathcal A(s)$  over all the devices in  $\mathcal M_w(s)$  can be calculated as:

\ifCLASSOPTIONonecolumn
         \begin{align}\label{eqn:Ireward}
        \bar{r}(s,a) = \sum_{R_k \in \mathcal M_w(s)}\bar{r}_k(s,a) = \sum_{R_k \in \mathcal X(a): v_{i,kl} \in \kappa(s)}  \delta_{k,l}(1-\epsilon_{i,k}).
      \end{align}
\else
        \begin{align}\label{eqn:Ireward}
        \bar{r}(s,a) &= \sum_{R_k \in \mathcal M_w(s)}\bar{r}_k(s,a) \nonumber \\
         &= \sum_{R_k \in \mathcal X(a): v_{i,kl} \in \kappa(s)}  \delta_{k,l}(1-\epsilon_{i,k}).
      \end{align}
\fi


\end{enumerate}

\subsection{MDP Solution Complexity} \label{complexity}
An MDP policy $\pi = [\pi(s)]$ is a mapping from state space   to action space  that specifies an  action to each of the states. Every policy is associated with a value function $V_{\pi} (s)$ that gives the expected cumulative reward at the end of  the deadline, when the system starts at state $s$ and follows policy $\pi$. It can be recursively expressed as \cite{puterman2009markov}:
\begin{equation}
V_{\pi} (s) = \bar{r}(s, a) + \sum_{s' \in \mathcal S(s,a)} \mathcal{P}_a(s,\acute{s}) V_{\pi} (s'), \;\;\;\;\;\forall s \in \mathcal S.
\end{equation}
Here, $\mathcal S(s,a)$ is the set of successor states to state $s$ when action $a =\kappa(s)$ is taken following policy $\pi(s)$.
The solution of a finite horizon MDP problem is an optimal policy $\pi^*(s)$ at state $s$ that maximizes  the expected cumulative  reward at the end of the finite number of time slots  and   can defined  as  \cite{puterman2009markov}:
\begin{equation}
\pi^*(s) = \arg\max_{a \in \mathcal A(s)}\{V_{\pi} (s)\},\;\;\;\; \forall s \in \mathcal S.
\end{equation}
The optimal policy  can be computed iteratively using the  backward induction algorithm (BIA). From the modeling perspective, BIA requires to define all state-action transition probabilities and rewards of all transitions.  From the computational perspective, it has complexity of $O(|\mathcal{S}|^2|\mathcal A|)$.
Based on the sizes of $\mathcal{S}$  and  $\mathcal{A}(s)$  described in our MDP formulation,   we conclude that finding the optimal policy using BIA is  computationally complex,  especially for systems with large numbers of  devices $M$  and packets $N$.
Therefore, in the following section, we design a low-complexity  IDNC algorithm that  can efficiently reduce the mean video  distortion before the deadline.

\section{Two-stage Maximal  Independent Set Selection Algorithm}\label{two-stage}
In this section, we  propose  a two-stage maximal independent set (TS-MIS) selection algorithm  that eliminates  the need for using BIA (a dynamic programming approach) and reduces both modeling and computational complexities. This is  a greedy approach since it selects an action in a given state without going through all the successor  states. However,   this approach follows the characteristics  of our  sequential decision making problem and   reduces the mean video distortion at the end of  the deadline. The main aspects of this approach are summarized as follows:

\begin{itemize}
\item We prioritize the critical devices  over the non-critical devices in making decisions. If a non-critical device is  ignored at the current time slot $t$, it is still possible to deliver all its missing packets in the remaining $Q-1$ time slots. On the other hand, a critical device already  has a larger number of missing packets compared to the remaining  time slots. Therefore, if a critical device is ignored at the current time slot $t$, it will receive a smaller subset of its missing packets at the end of the deadline.\footnotemark \footnotetext{Note that a non-critical device at time slot $t$ can become  a critical device at the successor  time slot $t+1$ and  have a high priority compared to other devices.}
\item To prioritize the critical devices, we partition   the  IDNC graph $\mathcal G$ into critical graph $\mathcal G_c$ and non-critical graph $\mathcal G_a$. The critical graph $\mathcal G_c$  includes  the vertices representing transmissions from all devices to the critical devices.  Similarly, the non-critical graph $\mathcal G_a$  includes the vertices representing transmissions from all devices to the non-critical  devices.

\item   It may not be possible to deliver all the missing packets to  the critical devices before the deadline due to their large numbers of missing packets.  Consequently, we select   a critical  maximal independent set $\kappa_{c}^*$ over critical  graph $\mathcal G_c$ that delivers the high importance packets to a subset of, or if possible, all critical devices.

\item  It is still  possible to deliver all the missing packets to the non-critical devices before the deadline due to  their  small numbers of missing packets. Consequently,  we select a non-critical  maximal independent set $\kappa_{a}^*$ over non-critical graph $\mathcal G_a$   that increases   the probability  of delivering all the missing packets to  all non-critical devices before the deadline. However,  $\kappa_{a}^*$ is selected  without violating the independent set  constraint (thus, prohibiting coding and transmission conflicts) for the targeted critical devices in $\kappa_c^*$.
\end{itemize}

\subsection{Maximal Independent Set Selection Algorithm  over  Critical Graph} \label{sec:CTT}
In this sub-section, we select a critical maximal independent set $\kappa_c^*$  over critical  graph $\mathcal G_c$ that minimizes the sum video distortion of all critical devices after the current time slot $t$. Let us define  $\mathcal X_c(\kappa_{c})$ as the set of targeted  critical devices  in  $\kappa_c$ and  $d_k^{(t+1)}(\kappa_c)$ as the expected individual video distortion of critical device $R_k \in \mathcal C(t)$  at time slot $t + 1$ due to selecting $\kappa_c$. This can be expressed  as:

\ifCLASSOPTIONonecolumn
  \begin{equation}
 d_k^{(t+1)}(\kappa_c) =
   \begin{cases}
    d_k^{(t)} & \text{if  $R_k \in \mathcal C(t) \setminus  \mathcal X_c(\kappa_c) $},  \\
    d_k^{(t)} - \delta_{k,l}(1-\epsilon_{i,k})  & \text{if  $R_k \in \mathcal X_c(\kappa_c): v_{i,kl} \in \kappa_c$}
   \end{cases}
 \end{equation}
\else
    \begin{align}
 &d_k^{(t+1)}(\kappa_c) \nonumber \\
  &  = \begin{cases}
    d_k^{(t)} & \text{if  $R_k \in \mathcal C(t) \setminus  \mathcal X_c(\kappa_c) $},  \\
    d_k^{(t)} - \delta_{k,l}(1-\epsilon_{i,k})  & \text{if  $R_k \in \mathcal X_c(\kappa_c): v_{i,kl} \in \kappa_c$}
   \end{cases}
 \end{align}
\fi

Here,  the first  term  represents the ignored critical device  for which the  distortion value  will remain unchanged from time slot $t$ to time slot $t + 1$. The second term represents the expected distortion reduction in the targeted critical device from time slot $t$ to time slot $t + 1$. Let $D^{(t+1)}(\kappa_c)$  be the sum of individual video distortion of all critical devices after  time slot $t$. We now express the expected sum video distortion of all critical devices after time slot $t$ as:

\ifCLASSOPTIONonecolumn
\begin{align} \label{eqn:distortion22}
\mathds{E}[D^{(t+1)}(\kappa_c)] &= \sum_{R_k \in \mathcal C(t)} \mathds{E}[ d_k^{(t+1)}(\kappa_c)] \nonumber \\
& =  \sum_{R_k \in \{\mathcal C(t) \setminus  \mathcal X_c(\kappa_c)\} } d_k^{(t)} +  \sum_{R_k \in   \mathcal X_c(\kappa_c)} d_k^{(t)} - \delta_{k,l}(1-\epsilon_{i,k}).
\end{align}
\else
\begin{align} \label{eqn:distortion22}
\mathds{E}[D^{(t+1)}(\kappa_c)] &= \sum_{R_k \in \mathcal C(t)} \mathds{E}[ d_k^{(t+1)}(\kappa_c)] \nonumber \\
& =  \sum_{R_k \in \{\mathcal C(t) \setminus  \mathcal X_c(\kappa_c)\} } d_k^{(t)} \nonumber \\
& \;\;\;\;\;\;\;\;\; +  \sum_{R_k \in   \mathcal X_c(\kappa_c)} d_k^{(t)} - \delta_{k,l}(1-\epsilon_{i,k})
\end{align}
\fi

We now formulate the problem of  minimizing the sum video distortion of all critical devices as a  critical maximal independent set $\kappa_c^*$  selection problem over  critical  graph  $\mathcal G_c$ such that:

\ifCLASSOPTIONonecolumn
\begin{align} \label{eqn:videoS}
 \kappa_c^* &= \arg\min_{\kappa_{c} \in \mathcal G_c} \mathds{E}[D^{(t+1)}(\kappa_{c})]  \nonumber  \\
& =  \arg\min_{\kappa_{c} \in \mathcal G_c} \{ \sum_{R_k \in \{\mathcal C(t) \setminus  \mathcal X_c(\kappa_{c})\} } d_k^{(t)} +  \sum_{R_k \in \mathcal X_c(\kappa_c)} d_k^{(t)} - \delta_{k,l}(1-\epsilon_{i,k})\} \\ \nonumber
& = \arg\max_{\kappa_{c} \in \mathcal G_c} \{ \sum_{R_k \in \mathcal X_c(\kappa_c) }   \delta_{k,l}(1-\epsilon_{i,k})\}.
\end{align}
\else
\begin{align} \label{eqn:videoS}
 \kappa_c^* &= \arg\min_{\kappa_{c} \in \mathcal G_c} \mathds{E}[D^{(t+1)}(\kappa_{c})]  \nonumber  \\
& =  \arg\min_{\kappa_{c} \in \mathcal G_c} \{ \sum_{R_k \in \{\mathcal C(t) \setminus  \mathcal X_c(\kappa_{c})\} } d_k^{(t)} \nonumber \\
& \;\;\;\; \;\;\;\; \;\;\;\;\;\;\;\;\;\;\;\;\;\;\;\;\;\;\;\;\;\; +  \sum_{R_k \in \mathcal X_c(\kappa_c)} d_k^{(t)} - \delta_{k,l}(1-\epsilon_{i,k})\}\nonumber \\
& = \arg\max_{\kappa_{c} \in \mathcal G_c} \{ \sum_{R_k \in \mathcal X_c(\kappa_c) }   \delta_{k,l}(1-\epsilon_{i,k})\}.
\end{align}
\fi

In other words, the problem of minimizing the sum video distortion  of all critical devices is equivalent to finding the maximum weighted  independent set in the critical graph $\mathcal G_c$. In this paper, we use the Bron-Kerbosch algorithm to find  $\kappa_c^*$  among all maximal independent sets in $\mathcal G_c$ \cite{bron1973algorithm}. The complexity of the Bron-Kerbosch algorithm of a graph with $|\mathcal V|$ vertices  is  $O(3^{|\mathcal V|/3})$.\footnotemark \footnotetext{To select a maximal independent set  with much lower computational  complexity,  a  greedy vertex search approach can be adopted  following  \cite{sorour2012completion}, which has  a tolerable  performance degradation.} In the following two sub-sections,  we first derive the probability that the individual completion times of all non-critical devices meet the deadline and  then select a non-critical   maximal independent set $\kappa_{a}^*$.

\subsection{Probability that the Individual Completion Time Meets Deadline}
At any given time slot $t$, we select a non-critical maximal independent set that  increases the probability   of  delivering all missing packets to all non-critical devices before the deadline. To select such an independent set, we  compute the  probability that the individual completion times of  all non-critical devices  meet the deadline. The computation of  this probability is simple since it is computed separately for each non-critical  device  and  does not take into account  the interdependence of devices' packet reception  captured in the GSM. In fact, we trade-off some accuracy in calculation for much more computational simplicity.

To derive  the  probability, we first consider a special  scenario with  a single non-critical device $R_k$ and assume that it is targeted with a new packet in each time slot.    The probability of individual completion time $T_{W_k}$ of device $R_k$ being equal to $W_k + x, x \in [0,1,...,Q-W_k]$  can be expressed using negative binomial distribution as:
\begin{equation} \label{eq:binomial}
  \mathds{P}[T_{W_k} = W_k + x] =  \binom{W_k + x -1}{x} (\bar{\epsilon}_k)^x(1-\bar{\epsilon}_k)^{W_k},
\end{equation}
where, $\bar{\epsilon}_k$ is the average  of the channel erasure probabilities  connecting device $R_k$ to other devices. In other words, $\bar{\epsilon}_k = \frac{\sum_{ R_i \in \mathcal I }\epsilon_{i,k}}{|\mathcal I|}$, where $\mathcal I = \{R_i|y_{i,k} \neq 0, R_i \neq R_k\}$. This average erasure probability  represents that  device $R_k$ can receive its missing packets from  any other  neighboring  device in  the remaining  time slots.  Consequently, the probability that the individual completion time $T_{W_k}$ of non-critical  device $R_k$ is less than or equal to the remaining $Q$  time slots  can be expressed as:
\begin{equation} \label{eq:metric}
\mathds{P}[T_{W_k} \leq Q] = \sum_{x = 0}^{Q - W_k}  \mathds{P}[T_{W_k} = W_k  + x].
\end{equation}
We now consider a scenario with  a set of non-critical devices $\mathcal A$ and assume that all  non-critical devices  are targeted with a new packet in
each   time slot. This is an ideal scenario and defines a lower bound on individual completion time of each non-critical device. Consequently, we can compute an upper bound on the probability that individual completion time of each non-critical device  meets the deadline. However, this ideal scenario will not occur in practice since the transmitting devices cannot benefit from their own transmissions and  the instant decodability constraint limits the number of targeted  devices in each time slot. We can still use this probability upper bound as a metric in designing our computationally simple IDNC algorithms.

With the aforementioned  ideal scenario, at any  D2D  time slot $t$, we can compute  the  upper bound on the probability that  individual completion times of all non-critical devices in $\mathcal A(t)$ are less than or equal to the remaining $Q$  time slots (denoted by $\hat{\mathds{P}}^{(t)} [T_A\leq Q]$) as:
\begin{equation} \label{eq:metric}
\hat{\mathds{P}}^{(t)}[T_A \leq Q] = \prod_{R_k \in \mathcal A(t)} \sum_{x=0}^{Q - W_k}
\mathds{P}[T_{W_k} = W_k + x].
\end{equation}

In the following sub-section, we use expression \eqref{eq:metric} as a metric of selecting a non-critical maximal independent set in each time slot.

\subsection{Maximal Independent Set Selection Algorithm over  Non-critical Graph}

Once a critical  maximal independent set $\kappa_c^*$ is selected over critical graph $\mathcal G_c$, there may exist vertices belonging to the non-critical devices  in  non-critical graph $\mathcal G_a$ that can form even a bigger maximal independent set. When the selected new vertices are non-adjacent  to all vertices in $\kappa_c^*$, the corresponding   non-critical devices are targeted without creating coding or transmission conflicts for the targeted critical devices in  $\kappa_c^*$. Therefore, we first extract  non-critical subgraph  $\mathcal G_a(\kappa_c^*)$ of vertices in $\mathcal G_a$  that are non-adjacent to all the vertices in $\kappa_c^*$  and then select non-critical maximal independent set  $\kappa_a^*$ over subgraph $\mathcal{G}_a(\kappa_c^*)$.

Let us define  $\mathcal X_a(\kappa_{a})$ as the set of targeted  non-critical devices  in  $\kappa_a$ and  $W_k^{(t+1)}(\kappa_a)$ as the expected number of missing packets at a non-critical  device $R_k \in \mathcal A(t)$  at time slot $t + 1$ due to selecting  $\kappa_a$. This can be expressed  as:
\ifCLASSOPTIONonecolumn
  \begin{equation} \label{eqn:wants}
 W_k^{(t+1)}(\kappa_a) =
   \begin{cases}
    W_k^{(t)} & \text{if  $R_k \in \mathcal A(t) \setminus  \mathcal X_a(\kappa_a) $},  \\
    (W_k^{(t)} -1)(1-\epsilon_{i,k}) + (W_k^{(t)})(\epsilon_{i,k})   & \text{if  $R_k \in \mathcal X_a(\kappa_a): v_{i,kl} \in \kappa_a$}
   \end{cases}
 \end{equation}
 \else
   \begin{align} \label{eqn:wants}
 W_k^{(t+1)}(\kappa_a) =
   \begin{cases}
    W_k^{(t)} & \text{}  \\
  \;\;\;\;\;\;\;\;\;\;\;\;\;\;\;\;\;\;\;\; \text{ if  $R_k \in \mathcal A(t) \setminus  \mathcal X_a(\kappa_a)$}, & \text{}  \\
   (W_k^{(t)} -1)(1-\epsilon_{i,k}) + (W_k^{(t)})(\epsilon_{i,k})   & \text{}\\
    \;\;\;\;\;\;\;\;\;\;\;\;\;\;\;\;\;\;\;\;\;\; \text{if  $R_k \in \mathcal X_a(\kappa_a): v_{i,kl} \in \kappa_a$}  & \text{}
   \end{cases}
 \end{align}
 \fi

Here, the first term represents the ignored non-critical device for which the number of missing packets will remain unchanged from time slot $t$ to time slot $t+1$. The second term represents the  targeted non-critical device for which the number of missing packets  can be either $W_k -1$  with the packet reception probability $(1-\epsilon_{i,k})$  or $W_k$ with the channel erasure probability  $\epsilon_{i,k}$. With $\kappa_a$ selection at time slot $t$, let $\hat{\mathds{P}}^{(t+1)} [T_A\leq Q-1]$ be the resulting   upper bound on the probability that  individual completion times of all non-critical devices in $\mathcal A(t)$, starting from the successor time slot $t + 1$, are less than or equal to the remaining $Q-1$  time slots. We can express   probability  $\hat{\mathds{P}}^{(t+1)} [T_A\leq Q-1]$  as:

\ifCLASSOPTIONonecolumn
\begin{align} \label{eqn:formulation3342}
      \hat{\mathds{P}}^{(t+1)} [T_A\leq Q-1] =  & \prod_{R_k \in   \mathcal{X}_{a}(\kappa_a)} \left( \mathds{P} [T_{W_k-1} \leq Q-1].(1-\epsilon_{i,k})+
             \mathds{P} [T_{W_k} \leq Q-1].(\epsilon_{i,k}) \right) \nonumber \\
              &  \times \prod_{R_k \in \mathcal{A} \setminus   \mathcal{X}_{a}(\kappa_a)}  \mathds{P} [T_{W_k} \leq Q-1]
\end{align}
\else
\begin{align} \label{eqn:formulation3342}
      \hat{\mathds{P}}^{(t+1)}& [T_A\leq Q-1] \nonumber \\
    & \;\;\;\; =   \prod_{R_k \in   \mathcal{X}_{a}(\kappa_a)} ( \mathds{P} [T_{W_k-1} \leq Q-1].(1-\epsilon_{i,k}) \nonumber \\
    & \;\;\;\;\;\;\;\;\;\;\;\;\;\;\;\;\;\;\;\;\;\;\;\;\;\;\;\;\; +  \mathds{P} [T_{W_k} \leq Q-1].(\epsilon_{i,k}) ) \nonumber \\
              & \;\;\;\;\;\;\; \times \prod_{R_k \in \mathcal{A} \setminus   \mathcal{X}_{a}(\kappa_a)}  \mathds{P} [T_{W_k} \leq Q-1]
\end{align}

\fi

In the first product, we   compute the probability that a targeted  non-critical device   receives  its $W_k-1$ or $W_k$ missing packets in the remaining $Q-1$ time slots. Moreover, in the  second product,  we compute the probability that an ignored  non-critical device  receives  its  $W_k$ missing packets in the remaining $Q-1$ time slots. We now formulate the problem of maximizing  probability  $\hat{\mathds{P}}^{(t+1)} [T_A\leq Q-1]$  as  a non-critical  maximal independent set $\kappa_a^*$ selection problem over non-critical  subgraph $\mathcal{G}_a(\kappa_c^*)$  such that:
\ifCLASSOPTIONonecolumn
\begin{equation} \label{eqn:formulation33}
\begin{split}
\kappa_a^*  &= \arg \;\max_{\kappa_a \in \mathcal G_a(\kappa_c^*) } \left\{\hat{\mathds{P}}^{(t+1)} [T_A \leq Q-1] \right\}  \\
             &= \arg \;\max_{\kappa_a \in \mathcal G_a(\kappa_c^*)} \{ \prod_{R_k \in
             \mathcal{X}_{a}(\kappa_a)} \left( \mathds{P} [T_{W_k-1} \leq Q-1].(1-\epsilon_{i,k})+
             \mathds{P} [T_{W_k} \leq Q-1].(\epsilon_{i,k}) \right)\\
              & \;\;\;\;\;\;\;\;\;\;\;\;\;\;\;\;\;\;\;\;\;\;\;\;\; \times \prod_{R_k \in \mathcal{A} \setminus \mathcal{X}_{a}(\kappa_a)}  \mathds{P}[T_{W_k} \leq Q-1]\}
\end{split}
\end{equation}
\else
\begin{align} \label{eqn:formulation33}
\kappa_a^*  &= \arg \;\max_{\kappa_a \in \mathcal G_a(\kappa_c^*) } \left\{\hat{\mathds{P}}^{(t+1)} [T_A \leq Q-1] \right\} \nonumber \\
             &= \arg \;\max_{\kappa_a \in \mathcal G_a(\kappa_c^*)} \{ \nonumber \\
             &\;\;\;\;\;\;\;\;\;\;\;\;\;  \prod_{R_k \in
             \mathcal{X}_{a}(\kappa_a)} ( \mathds{P} [T_{W_k-1} \leq Q-1].(1-\epsilon_{i,k}) \nonumber  \\
             & \;\;\;\;\;\;\;\;\;\;\;\;\;\;\;\;\;\;\;\;\;\;\;\;\;\;\;\;\;\;\;  + \mathds{P} [T_{W_k} \leq Q-1].(\epsilon_{i,k}) ) \nonumber\\
              & \;\;\;\;\;\;\;\;\; \times \prod_{R_k \in \mathcal{A} \setminus \mathcal{X}_{a}(\kappa_a)}  \mathds{P}[T_{W_k} \leq Q-1]\}
\end{align}
\fi
In other words, the problem of maximizing probability $\hat{\mathds{P}}^{(t+1)} [T_A\leq Q-1]$    is equivalent to finding all maximal independent sets  in the non-critical subgraph $\mathcal G_a(\kappa_c^*)$, and selecting the maximal independent set among them that results in the maximum probability $\hat{\mathds{P}}^{(t+1)} [T_A\leq Q-1]$.
Similar to Section \ref{sec:CTT}, we use the Bron-Kerbosch algorithm to find $\kappa_a^*$ among all  maximal independent sets in $\mathcal G_a(\kappa_c^*)$.

The  final  maximal independent set $\kappa^*$ is the union of two maximal independent sets $\kappa_c^*$ and  $\kappa_a^*$  (i.e., $\kappa^* = \{\kappa_c^* \cup
\kappa_a^* \}$). All the  vertices  in $\kappa^*$ determines  a set of transmitting devices. Each of the selected transmitting devices forms a coded packet by XORing the source packets identified by the vertices in $\kappa^*$ representing transmission from that device.  The proposed two-stage maximal independent set (TS-MIS) selection  algorithm is summarized in Algorithm \ref{alg:LGS}.

\IncMargin{1em}
\begin{algorithm}[t]
\textbf{Construct}  IDNC graph $\mathcal G$ according to all LSMs $\mathbf F_{i}, \forall R_i \in \mathcal M$\;
Partition $\mathcal G$ into $\mathcal G_c$ and $\mathcal G_a$ according to the critical   and the non-critical devices\;
Initialize $\kappa_c^* = \varnothing$ and $\kappa_a^* = \varnothing$\;
 \If{$\mathcal G_c \neq \varnothing$}{
       Select $ \kappa_c^*=\arg\max_{\kappa_{c} \in \mathcal G_c} \left\{ \sum_{R_k \in \mathcal X_c }   \delta_{k,l}(1-\epsilon_{i,k}) \right\}$ \;
}
Update  subgraph  $\mathcal G_a(\kappa_c^*)$\;
 \If{$\mathcal G_a(\kappa_c^*) \neq \varnothing $}{

       Select $\kappa_a^* = \arg \;\max_{\kappa_a \in \mathcal G_a(\kappa_c^*) } \left\{\hat{\mathds{P}}^{(t+1)} [T_A \leq Q-1] \right\} $\;

}
\textbf{Set} $\kappa^* \leftarrow \kappa_c^* \cup \kappa_a^*$\;
\caption{Two-Stage Maximal Independent Set (TS-MIS) Selection Algorithm} \label{alg:LGS}
\end{algorithm}\DecMargin{1em}

\section{Calculations for Packet Importance of a  Real Video Sequence}\label{sec:distortion}

In this section, we first discuss the H.264/SVC video test sequence used in this paper and then provide details about  the  calculations for individual packet importance.  We use a standard video sequence, \emph{Soccer} \cite{test2014}. This sequence is in common intermediate format (CIF, i.e., $352 \times 288$) and has 300 frames with 30 frames per second (fps). We  encode the sequence using  the JSVM 9.19.14 version of H.264/SVC codec \cite{software2011,schwarz2007overview} while considering  the temporal scalability of the video sequence.\footnotemark \footnotetext{Note that our proposed IDNC framework is general and can be applied to a single layer H.264/AVC video sequence  considered in \cite{seferoglu2009video,keshtkarjahromi2015content}.}  The  size of each group of pictures (GOP) is $8$ frames, which results in $38$ GOPs for the  video sequence. As shown in Fig. \ref{fig:GOP}, each GOP consists of a sequence of I, P and B frames that are encoded into four video layers. We use the identical shade to represent the frames of the same video layer and the darker shades to represent the more important video layers. Moreover, we use arrows to  illustrate  the dependency between frames in a GOP. The GOP shown in Fig. \ref{fig:GOP} is a closed GOP, where the decoding of frames inside the GOP is independent of frames outside the GOP  \cite{esmaeilzadeh2014inter}.

We use  $1500$ bytes as the packet length. This  is the largest allowed packet over Ethernet. We allocate 1400 bytes for video information and the remaining 100 bytes for all the header information. Given  the encoded I frame (i.e., the first layer) composed of  $\sigma$ bytes, the required number of packets for this  frame and layer can be calculated as $\lceil\frac{\sigma}{1400}\rceil$. Here, the  ceiling function $\lceil.\rceil$  represents the additional padding bits that  are inserted into the  last packet of the layer to make it $1500$ bytes. The average number of packets in the first, second, third and fourth video layers over $38$ GOPs are $8.35, 3.11, 3.29$ and $3.43,$ respectively. This means   on average $8.35$ packets are required to decode  the first layer, which  consists of a single I frame. This frame is discarded at the devices if all the packets of this frame are not received before the deadline. For a GOP of interest, given that the number of frames per GOP is $8$, the video frame rate is 30 frames per second, the transmission rate is $\lambda$ bits per second  and a packet length is $1500 \times 8$ bits, the allowable number of total  time slots  for a GOP is fixed and can be computed as: $\frac{8\lambda}{1500 \times 8 \times 30}$.

\ifCLASSOPTIONonecolumn
\begin{figure}[t]
        \centering
        \includegraphics[width=8cm,height=5cm]{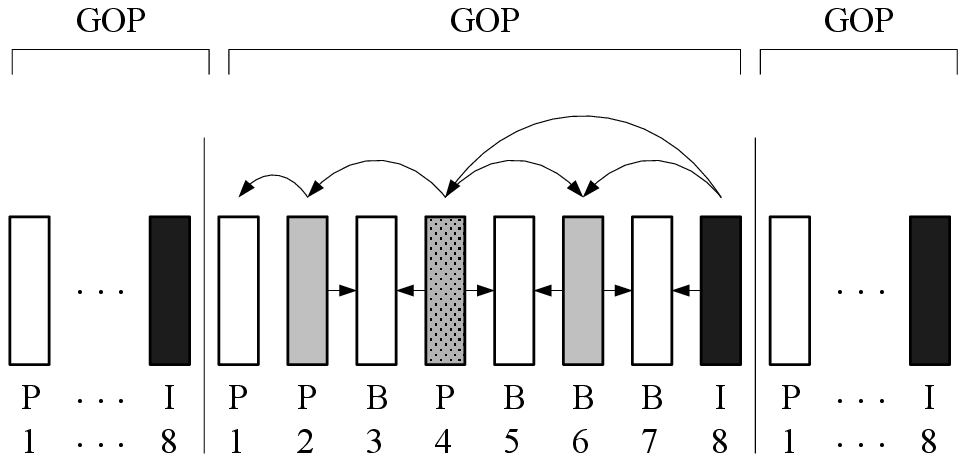}
        \caption{A closed GOP with 4 layers and 8 frames (a sequence of I, P and B frames).} \label{fig:GOP}
\end{figure}
\else
\begin{figure}[t]
        \centering
        \includegraphics[width=\columnwidth]{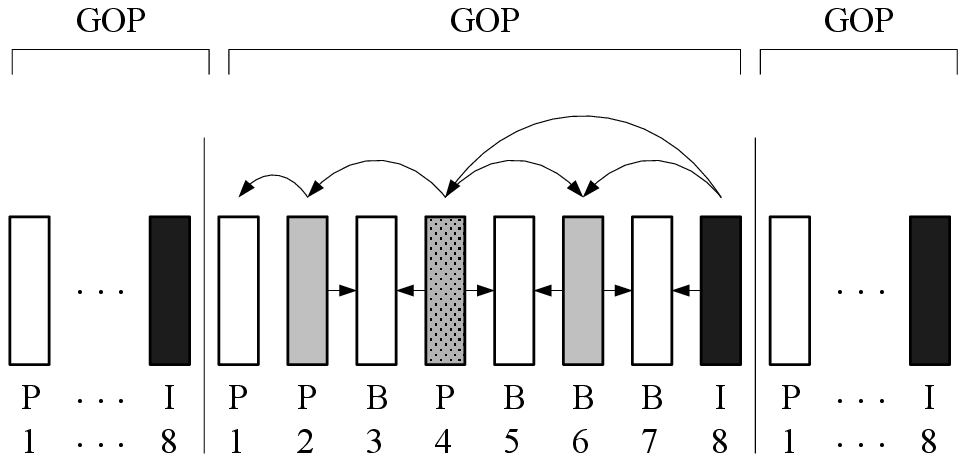}
        \caption{A closed GOP with 4 layers and 8 frames (a sequence of I, P and B frames).} \label{fig:GOP}
\end{figure}
\fi

In this paper, we  use the average \emph{peak-signal-to-noise ratio} (PSNR)  as the performance metric for the  video quality of our encoded video sequence \emph{Soccer}. Similar to the work in \cite{esmaeilzadeh2014inter},  we obtain $\alpha_{f_i, f_{j}} $ for $ 1 \leq f_i, f_j \leq 300$, which represents the PSNR if uncompressed $f_i$ frame is replaced by compressed $f_j$  frame. We calculate the average PSNR of each GOP, if the first $\ell$ layers of four video layers  are docodable $(0 \leq \ell \leq 4)$.\footnotemark \footnotetext{Note that the $\ell$-th  layer of a scalable video can be decoded  only if all packets in the first $\ell$ layers  are received  before the deadline.}  Moreover, the   frames  of the undecodable layers of the current GOP are replaced by the nearest frames in time of decodable layers of the current GOP or the previous GOP. This results in  concealing the errors in the video sequence. For example,  the average PSNR of the second GOP can be  calculated as:
\begin{align}
\bar{\alpha}_2 = \frac{ \sum_{f_i \in \mathcal B}\alpha_{f_i, f_i} +   \sum_{f_i \notin \mathcal B}\alpha_{f_i, f_{j}} }{8}
\end{align}
where,  $\mathcal B$ is the set of  frames  of the decodable layers of  the second GOP.

\begin{examples} \label{ex:error}
Let us  consider  the  GOP   shown in Fig. \ref{fig:GOP}. We  assume that the  fourth layer  of the second GOP is lost due to missing a packet of that layer at the end of  the deadline. The resulting error concealment is shown in Fig. \ref{fig:GOPerror} and  the resulting  average PSNR can be computed as:
\ifCLASSOPTIONonecolumn
\begin{align} \label{eq:conceal}
\bar{\alpha}_2  = \frac{\alpha_{f_1,f_{2}} + \alpha_{f_{2},f_{2}} + \alpha_{f_{3},f_{4}} + \alpha_{f_{4},f_{4}} + \alpha_{f_{5},f_{6}} + \alpha_{f_{6},f_{6}} + \alpha_{f_{7},f_{8}} + \alpha_{f_{8},f_{8}} }{8}
\end{align}
\else
\begin{align} \label{eq:conceal}
\bar{\alpha}_2  = & (\alpha_{f_1,f_{2}} + \alpha_{f_{2},f_{2}} + \alpha_{f_{3},f_{4}} + \alpha_{f_{4},f_{4}} \nonumber \\
&\;\;\; + \alpha_{f_{5},f_{6}} + \alpha_{f_{6},f_{6}} + \alpha_{f_{7},f_{8}} + \alpha_{f_{8},f_{8}})/8
\end{align}
\fi
\end{examples}

\begin{remarks}
(PSNR without Error) The average PSNR of the encoded  Soccer sequence  over $38$ GOPs is  $35.64$ decibel (dB)    if there is no error in the  sequence.
\end{remarks}

\ifCLASSOPTIONonecolumn
\begin{figure}[t]
        \centering
        \includegraphics[width=8cm,height=5cm]{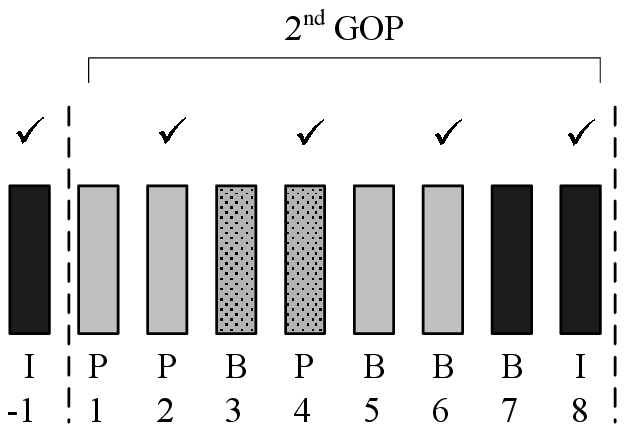}
        \caption{The nearest decoded frames are used  to conceal the loss
of undecoded frames.} \label{fig:GOPerror}
\end{figure}
\else
\begin{figure}[t]
        \centering
        \includegraphics[width=\columnwidth]{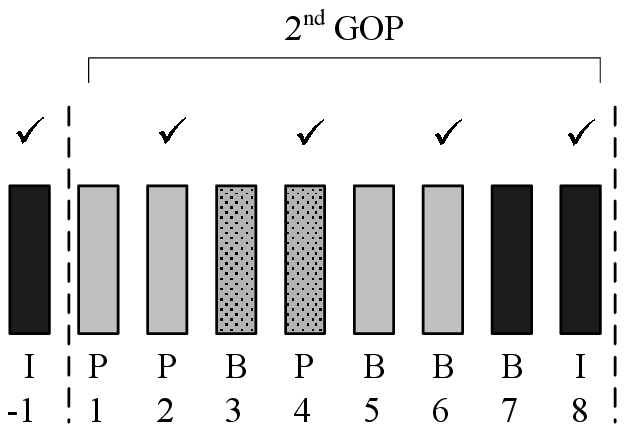}
        \caption{The nearest decoded frames are used  to conceal the loss
of undecoded frames.} \label{fig:GOPerror}
\end{figure}
\fi


\section{Simulation Results} \label{results}

In this section, we  present the simulation results comparing the performance of the BIA that solves the formulated MDP problem  and the  TS-MIS algorithm to the following algorithms.
\begin{itemize}
\item   `Fully Connected Distortion (FCD)'  algorithm \cite{keshtkarjahromi2015content} that considers a fully connected network  and uses IDNC  to minimize the mean video distortion in each time slot. This algorithm first determines the importance of individual packet  according to its contribution to the overall video quality. It then selects  a  transmitting device  and its XOR packet combination that  minimizes the mean video distortion after the current time slot.
\item `Partially Connected Blind (PCB)' algorithm  \cite{douik2014delay} that  considers a partially connected network  and uses IDNC  to serve the maximum number of devices with any new packet in each   time slot.  This  algorithm selects a set of transmitting devices and their XOR packet combinations  while  ignoring the hard  deadline  and the unequal importance of video packets. This  problem was addressed  in \cite{aboutorab2013instantly} for a fully connected D2D  network and  in \cite{le2013instantly} for a PMP network.
\end{itemize}

We first  consider a line network with $M = 4$ devices described in \eqref{eqn:localIDNC}  and encode  four video layers  of   \emph{Soccer}  video sequence   into four different    packets, i.e., $N=4$. As discussed in Section \ref{complexity}, the modelling and computational  complexities of the BIA  scale with the size of the state space $|\mathcal{S}|$, which is  $O(2^{16})$  even for $M = N = 4$.  Moreover, as discussed in Section \ref{tools}, the central station uses the initial $N $ time slots. Due to erasures in long-range wireless channels, at the beginning of the D2D phase,  each device  holds    between  $45\%$ and $55\%$ of $N$ packets in all scenarios. Note that these percentages of initial received  packets are arbitrary and  reflect the  erasures in long-range wireless channels.

\begin{definitions}
(Mean PSNR Calculation) The mean PSNR  is calculated by taking  average of the received  PSNR  at all  $M$ devices at the end of the deadline.
\end{definitions}

\ifCLASSOPTIONonecolumn
    \begin{figure}[t]
        \centering
        \includegraphics[width=12cm,height=8.5cm]{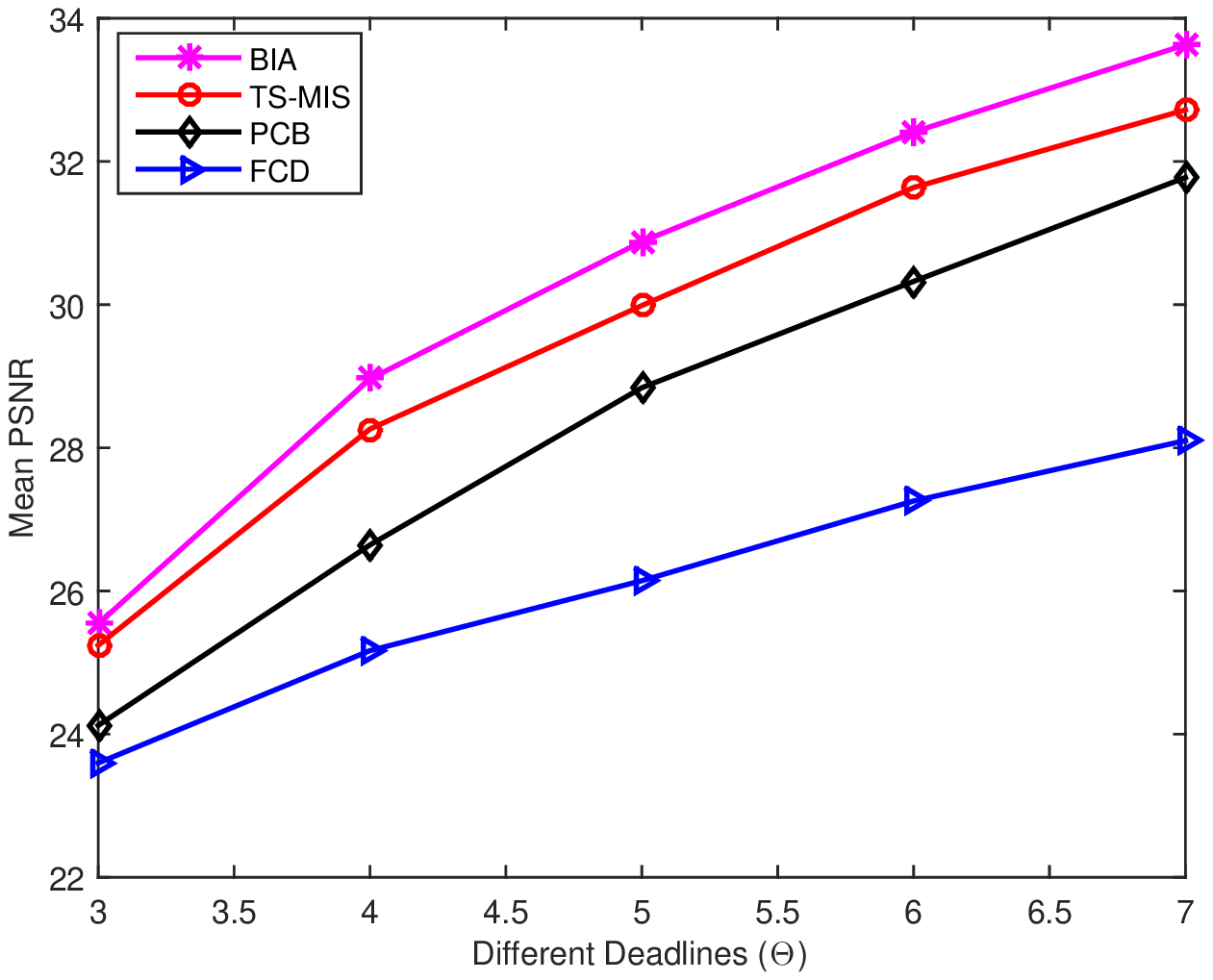}
        \caption{Mean PSNR versus different  deadlines $\Theta$.}\label{fig:mdp}
\end{figure}
\else
      \begin{figure}[t]
        \centering
        \includegraphics[width=\columnwidth]{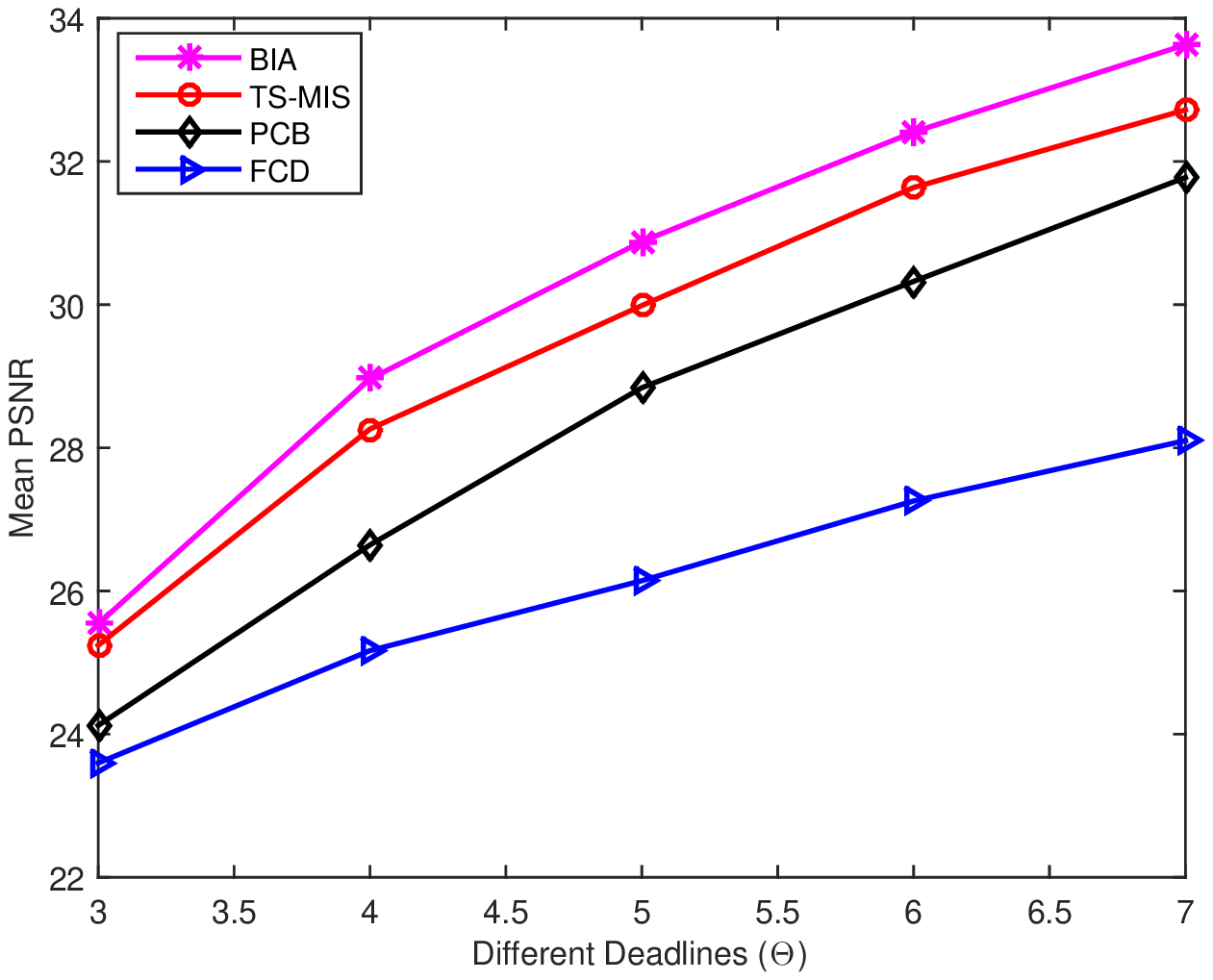}
        \caption{Mean PSNR versus different  deadlines $\Theta$.}\label{fig:mdp}
       \end{figure}

\fi

Fig. \ref{fig:mdp}  shows the mean PSNR achieved by different algorithms against the different number of allowable D2D time slots $\Theta$ (i.e., different deadlines). From this figure,  we can  see that our proposed  BIA and TS-MIS algorithms quickly increase the received PSNR at the devices  with increasing the deadlines. Indeed, both BIA and TS-MIS  algorithms  use the new  IDNC graph to make coding and transmission conflict-free decisions and exploit the characteristics of a real-time video sequence. This figure also shows that the performance of the  FCD  and PCB algorithms considerably deviates from  the BIA and TS-MIS algorithms.  FCD algorithm selects a single transmitting device and its  packet combination  without exploiting  the possibility of simultaneous transmissions  from multiple devices. Moreover, FCD algorithm does not capture the aspects of   the hard deadline and the channel erasures in making decisions. On the other hand, PCB  algorithm exploits the possibility of simultaneous transmissions from  multiple devices, but targets a large number of devices with any new packet in each time slot.

Fig. \ref{fig:hist} shows the histogram obtained by different algorithms for the same line network (for $M = N = 4$ and $\Theta = 7$). This histogram illustrates the percentage of received  PSNR     before the deadline at individual devices separately. From this histogram, we can see that  all devices receive an acceptable  video quality at the end of the deadline (i.e., $\Theta = 7$ D2D time slots). Moreover, devices $R_2$ and $R_3$ experience a slightly  better video quality compared to devices  $R_1$ and $R_4$ since these are the intermediate devices in the line network shown in Fig. \ref{fig:line}.

\ifCLASSOPTIONonecolumn
    \begin{figure}[t]
        \centering
        \includegraphics[width=12cm,height=8.5cm]{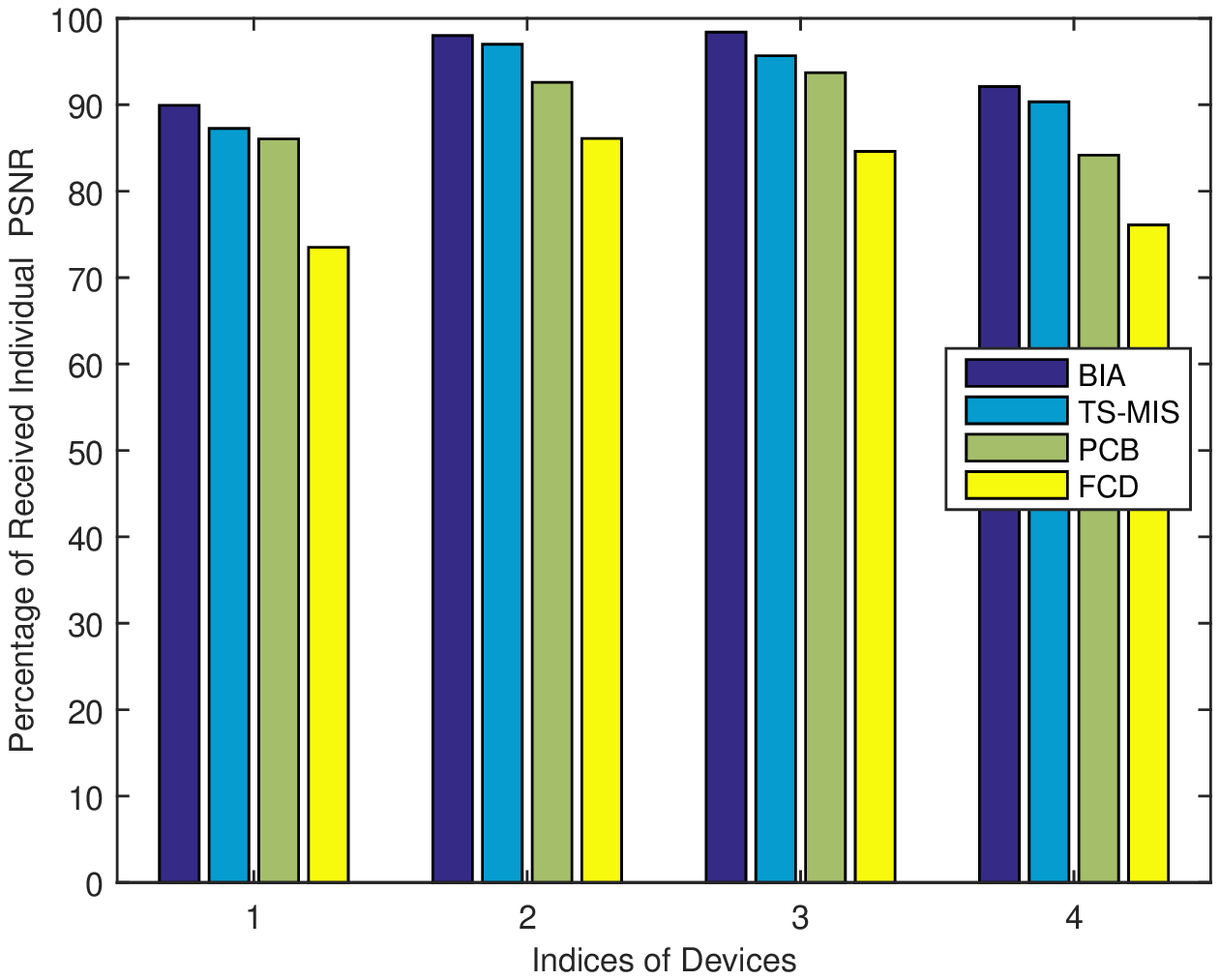}
        \caption{Histogram showing the percentage of received  PSNR at individual devices before the deadline.}\label{fig:hist}
\end{figure}
\else
      \begin{figure}[t]
        \centering
        \includegraphics[width=\columnwidth]{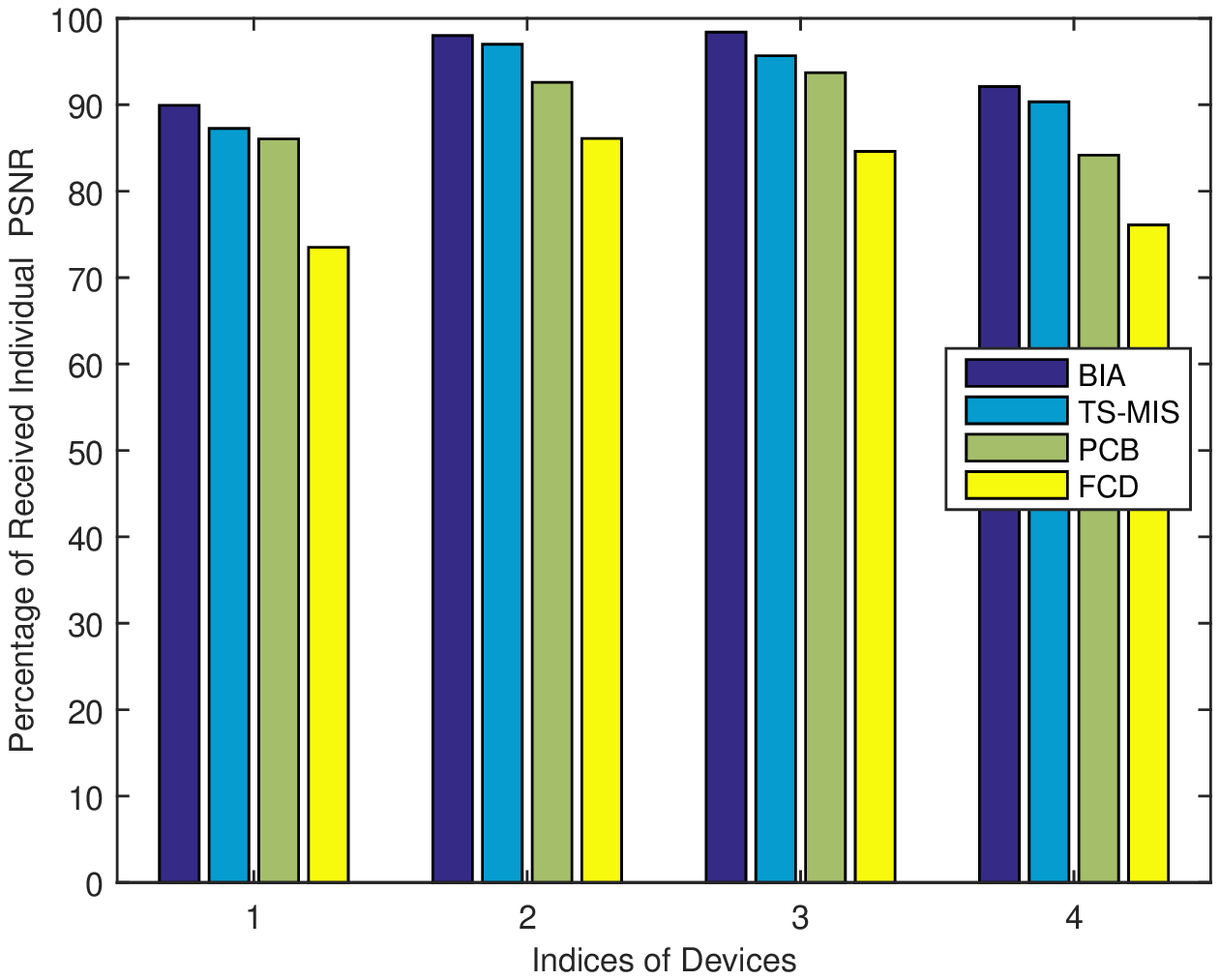}
        \caption{Histogram showing the percentage of received  PSNR at individual devices before the deadline.}\label{fig:hist}
       \end{figure}

\fi


Having shown the performance of the BIA and TS-MIS algorithms for a simple line network, we now consider more general partially connected  networks and show the performance of the  TS-MIS algorithm. We  use the \emph{Soccer} video sequence discussed in Section \ref{sec:distortion}, where the packet length is  1500 bytes and  each video layer is encoded into multiple packets. In SCM $\mathbf Y$, if  a pair of  devices are directly connected,   the packet  reception probability over the  channel is in the range $[0.65, 0.9]$. We compute the average  connectivity index in the network as  $\bar{y} = \frac{\sum_{(R_i,R_k)}y_{i,k}}{M \times M}$, which represents the average  packet reception probability over  all short-range channels. In the  case of a fully connected network, the average connectivity index is $\bar{y} = 0.8$.

\ifCLASSOPTIONonecolumn
\begin{figure}[t]
        \centering
        \includegraphics[width=12cm,height=8.5cm]{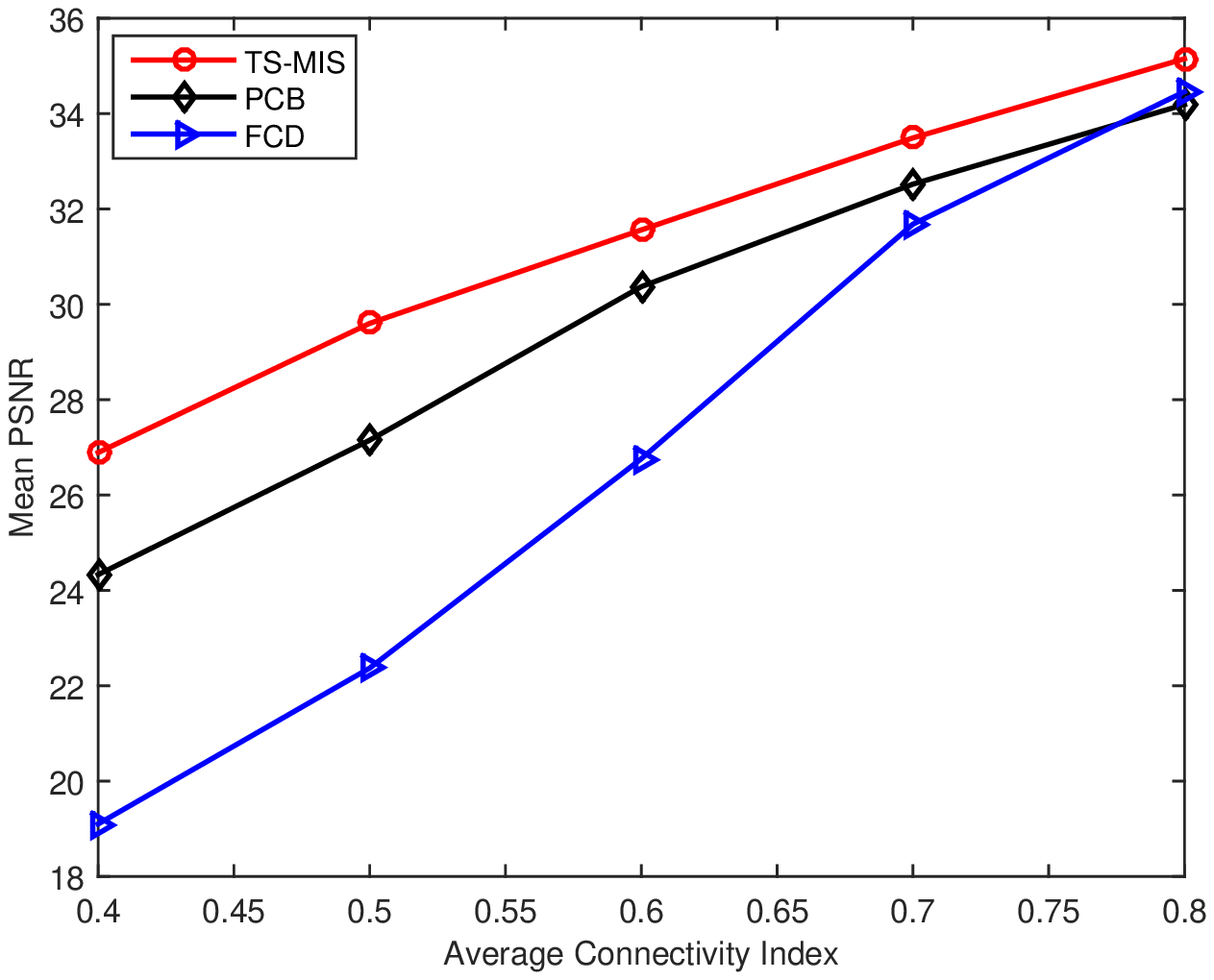}
        \caption{Mean PSNR versus different  average  connectivity indices $\bar{y}$.} \label{fig:connect}
\end{figure}
\else
    \begin{figure}[t]
        \centering
        \includegraphics[width=\columnwidth]{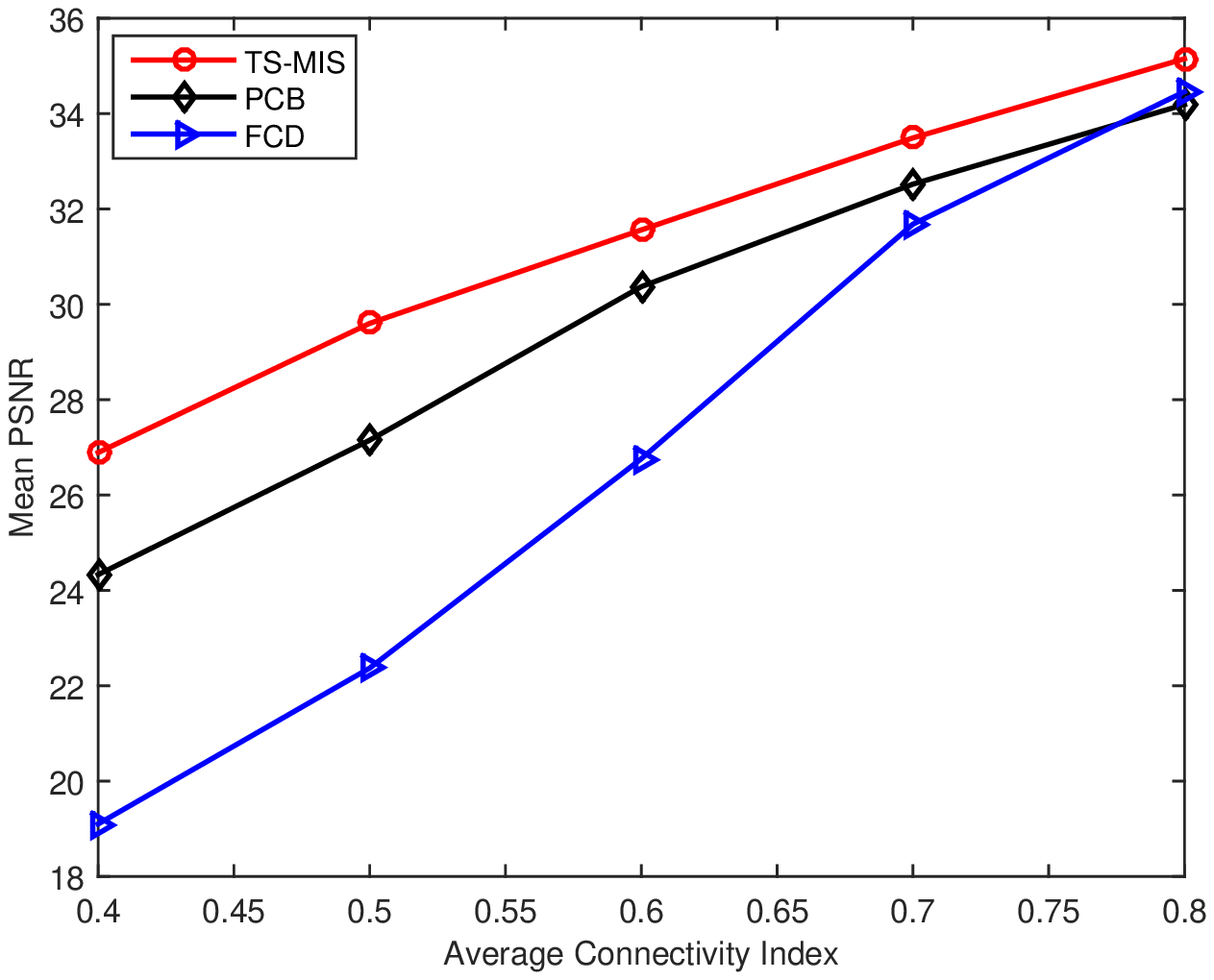}
        \caption{Mean PSNR versus different  average  connectivity indices $\bar{y}$.} \label{fig:connect}
        \end{figure}
\fi

\ifCLASSOPTIONonecolumn
\begin{figure}[t]
        \centering
        \includegraphics[width=12cm,height=8.5cm]{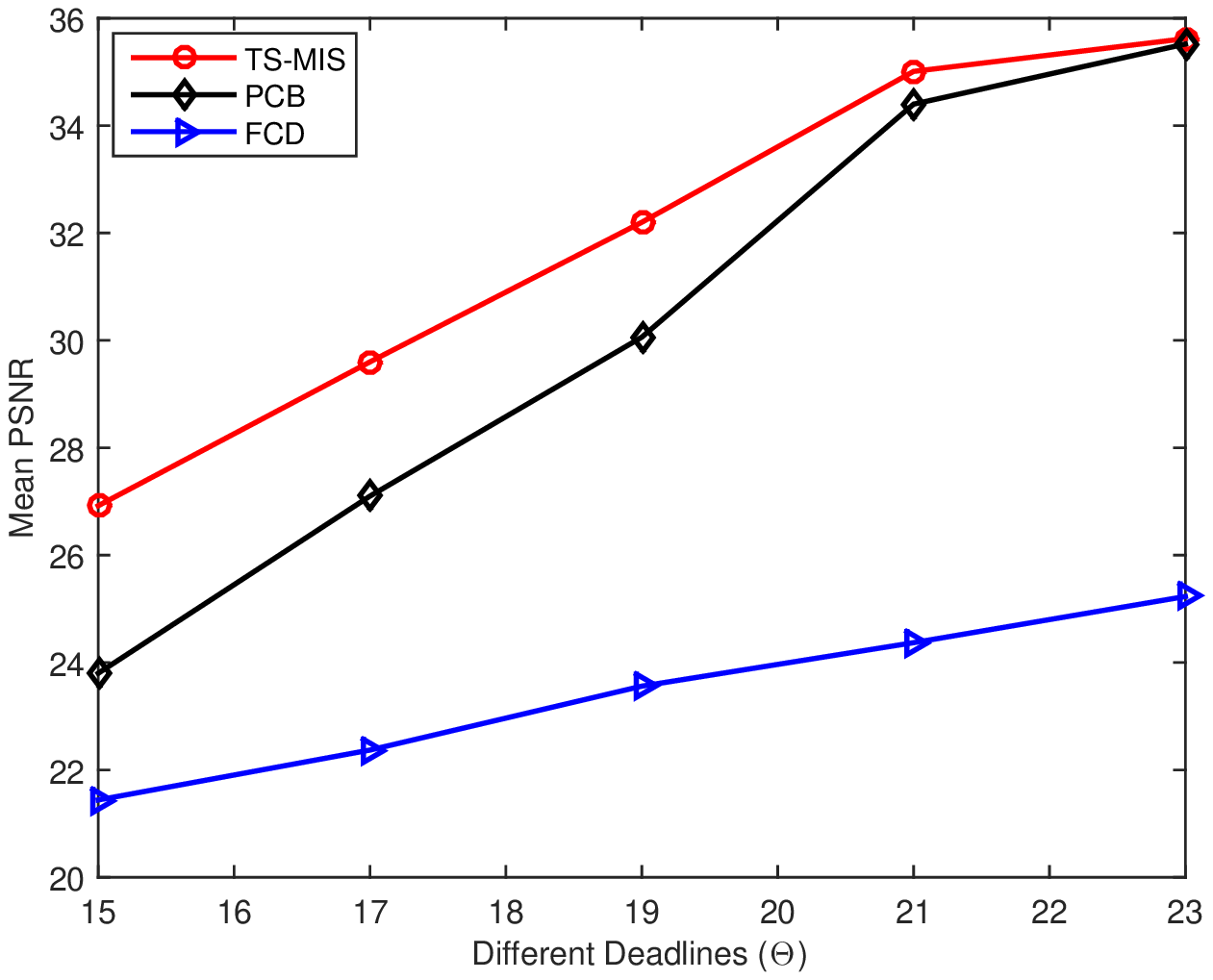}
        \caption{Mean PSNR versus different number of allowable time slots  $\Theta$} \label{fig:deadline}
\end{figure}
\else
\begin{figure}[t]
        \centering
        \includegraphics[width=\columnwidth]{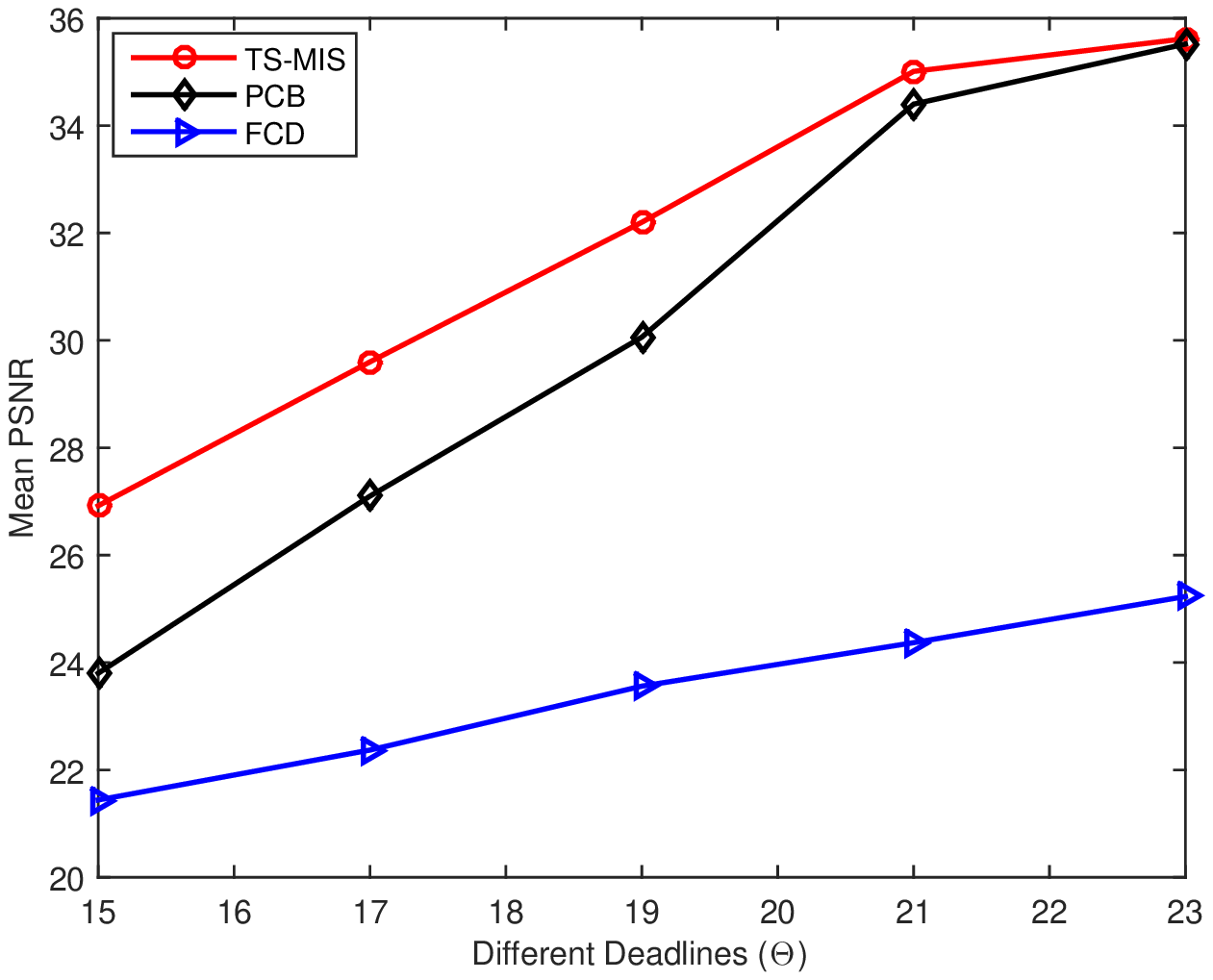}
        \caption{Mean PSNR versus different number of allowable time slots  $\Theta$} \label{fig:deadline}
\end{figure}
\fi

Fig. \ref{fig:connect}  shows the mean PSNR achieved by different algorithms against different average connectivity indices $\bar{y}$  (for $M = 15$ devices and  $\Theta =17$ D2D time slots). From this figure, we can see that our proposed TS-MIS algorithm outperforms the FCD algorithm in all cases,  even in the case of a fully connected network, i.e.,  $\bar{y} = 0.8$. In fact, our proposed TS-MIS algorithm adopts a decision that not necessarily minimizes the mean video distortion after the current time slot but  rather  reduces the mean video distortion at the end of the deadline. Moreover, the decisions of the  TS-MIS algorithm are adaptive  to the number of remaining time slots. In particular, when the number of remaining time slots is large and all devices are non-critical devices, generally as  in the case of the beginning of the D2D phase, the algorithm  increases  the probability of delivering all the packets to all devices. On the other hand, when the number of remaining time slots is small and all devices are critical devices, generally as in the case of the  end of the D2D phase,  the algorithm  minimizes the mean video distortion after the current time slot. Finally, the algorithm mixes both decisions when some devices are critical devices  and some are non-critical devices, in which case it prioritizes the critical devices since they  will receive one less packet with each ignored time slot  at the end of  the deadline.  From this figure, we can also see that the performance of the  PCB algorithm considerably deviates from the  TS-MIS algorithm  since   PCB algorithm does not  address the hard deadline  for the  high importance video packets.

\ifCLASSOPTIONonecolumn
\begin{figure}[t]
        \centering
        \includegraphics[width=12cm,height=8.5cm]{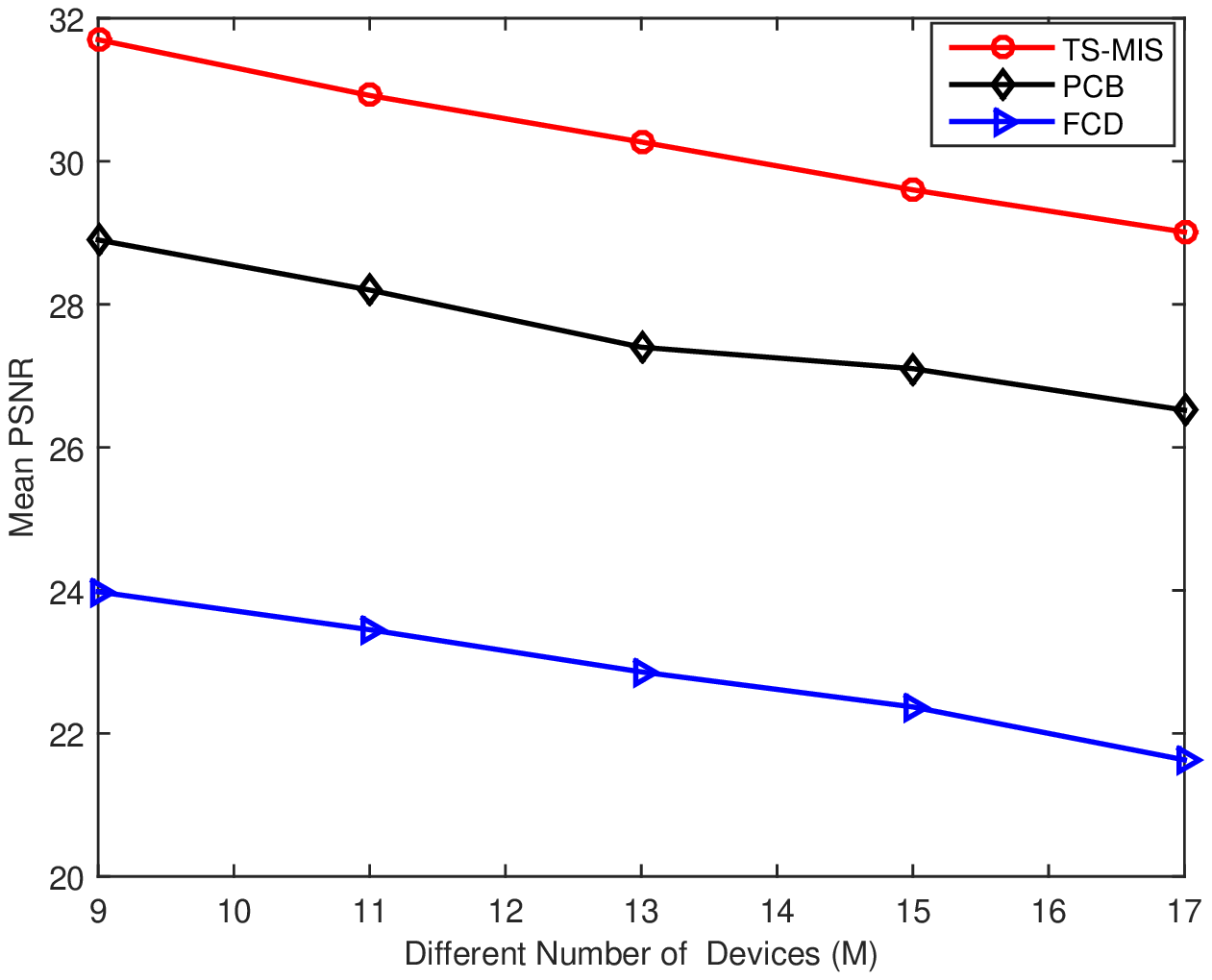}
        \caption{Mean PSNR versus different  number of devices $M$}\label{fig:rx}
\end{figure}
\else
\begin{figure}[t]
        \centering
        \includegraphics[width=\columnwidth]{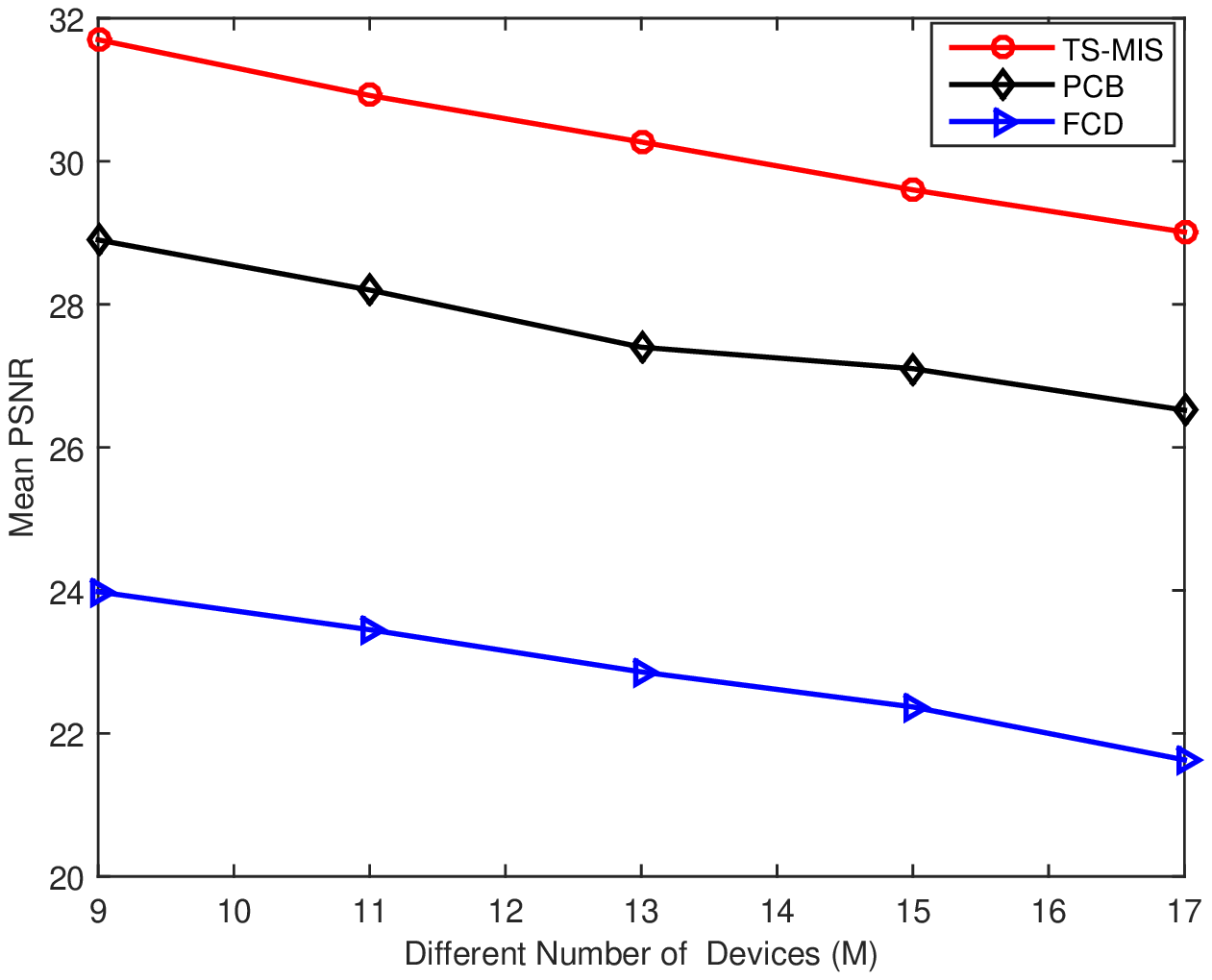}
        \caption{Mean PSNR versus different  number of devices $M$}\label{fig:rx}
\end{figure}
\fi

Fig. \ref{fig:deadline} and Fig. \ref{fig:rx}  show the mean PSNR achieved by different algorithms against  different deadlines $\Theta$ (for  $\bar{y} = 0.5$ average connectivity index and  $M = 15$ devices)   and different  number of devices $M$  (for  $\bar{y} = 0.5$ average connectivity index and $\Theta = 17$ D2D time slots), respectively. As expected, our proposed TS-MIS algorithm  outperforms the FCD and PCB algorithms in all scenarios. In fact, our proposed TS-MIS algorithm makes decisions by taking  into account the unequal importance of video packets, hard deadline, erasures of wireless channels, coding and transmission conflicts. Note that we have  used another video sequence \emph{Foreman} in the simulations  and observed the  similar  results as in the case of \emph{Soccer}.


%
\section{Conclusion} \label{conclusion}
In this paper, we  developed an efficient  IDNC framework for distributing a real-time video sequence to a group of cooperative  wireless  devices in a partially connected network. In particular, we  introduced  a novel IDNC graph that represents all feasible coding and transmission conflict-free decisions in one unified framework. Using the new  IDNC graph and   the characteristics of a real-time  video sequence, we   formulated the problem of minimizing the mean video distortion before the deadline as a finite horizon MDP problem. Since solving the formulated MDP problem was computationally complex, we further  designed  a TS-MIS selection algorithm that efficiently solves the problem with much lower complexity. Simulation results over a real video sequence showed that our proposed IDNC algorithms improve the received video quality compared to existing IDNC algorithms. Future research direction is to extend our proposed  IDNC framework  to a non-cooperative   system, where the devices are selfish and  pursue to minimize  their individual  video distortions before the deadline.

\vspace{-1mm}
\bibliographystyle{IEEEtran}
\bibliography{ref}

\end{document}